\documentclass[12pt]{elsart}
\usepackage{graphicx}
\usepackage{amsmath}
\usepackage{lscape}

\usepackage{colordvi}

\def\om{\omega}
\def\be{\begin{eqnarray}}
\def\ee{\end{eqnarray}}

\def\prt{\partial}

\newcommand{\lsim}{\stackrel{\scriptstyle <}{\phantom{}_{\sim}}}
\newcommand{\gsim}{\stackrel{\scriptstyle >}{\phantom{}_{\sim}}}
\def\rmF{{\rm F}}

\begin{document}

\begin{frontmatter}
\title{Relativistic Mean-Field Models with Scaled Hadron Masses and
Couplings: Hyperons and Maximum Neutron Star Mass}
 \author[MEPHI]{K. A.~Maslov},
 \author[UMB]{E. E.~Kolomeitsev} \and
 \author[MEPHI]{D. N.~Voskresensky}
\address[MEPHI]{ National Research Nuclear  University (MEPhI), 115409 Moscow, Russia}
\address[UMB]{Matej Bel  University, SK-97401 Banska Bystrica, Slovakia}
\begin{abstract}
An equation of state of cold nuclear matter with an arbitrary
isotopic composition is studied within a relativistic mean-field
approach with hadron masses and coupling constants depending self-consistently on
the scalar mean-field. All hadron masses decrease universally with the scalar field growth, whereas meson-nucleon coupling constants can vary differently. More specifically we focus on two modifications of the KVOR model studied previously. One extension of the model (KVORcut) demonstrates that the equation of state stiffens if the increase of the scalar-field magnitude with the density is bounded from above at some value for baryon densities above the saturation nuclear density. This can be realized if the nucleon vector-meson coupling constant changes rapidly as a function of the scalar field slightly above the desired value.
The other version of the model (MKVOR) utilizes a smaller value of the nucleon
effective mass at the nuclear saturation density and a saturation of the scalar field in the isospin asymmetric matter induced by a strong variation of the nucleon isovector-meson coupling constant as function of the scalar field. A possibility of hyperonization of the matter in neutron star interiors is incorporated. Our equations of state fulfill majority of known empirical constraints including the pressure-density constraint from heavy-ion collisions, direct Urca constraint, gravitational-baryon mass constraint for the pulsar J0737-3039B, and the constraint on the maximum mass of the neutron stars.
\end{abstract}
\end{frontmatter}
\tableofcontents

\section{Introduction}

It is convenient to describe the equation of state (EoS) of the baryon matter at densities relevant for heavy-ion collisions and  neutron stars  in terms of relativistic mean-field (RMF) models, cf.~\cite{Durr,SerotWalecka,Glendenning} and references therein. The original Walecka model~\cite{Walecka1974} included interaction of nucleons with mean fields of scalar ($\sigma$) and vector ($\omega$) mesons. Then  an isovector ($\rho$) meson field was
included to fit better the symmetry energy coefficient. The coupling constants of the meson fields with nucleons were fitted to describe the nuclear saturation, namely, the equilibrium density and the corresponding binding and symmetry energies. Next, a $\sigma$ field-dependent potential $U(\sigma)=b\sigma^3/3+c\sigma^4/4$ with two extra parameters $b$
and $c$ was added~\cite{Boguta77}, which allowed to fit values of the nucleon effective mass and the incompressibility coefficient. In all these models the nucleon mass is considered as a function of the $\sigma$ meson field, whereas the masses of the scalar, vector and isovector meson fields are treated as constants. Coupling constants are also assumed to be density and field independent quantities. At present, there exists a vast number of generalizations and modifications of the RMF models. They differ by new terms in an
effective Lagrangian related to new fields and their interactions~\cite{Toki,Sharma:2008jza}.
Further, RMF models with density dependent nucleon-meson coupling constants were developed, cf. Refs.~\cite{Long,Fuchs,Typel,Hofmann,Niksirc,Lalazisis,Gaitanos}.
A new  density functional DD-MEN based to a large extent on microscopic ab initio calculations in
nuclear matter was presented in~\cite{RocaMaza:2011qe}, which
includes nucleons interacting with $\sigma$, $\omega$, $\rho$
and scalar-isovector $\delta$  meson mean fields with density dependent
meson-nucleon coupling constants. For a detailed description  of the models with density dependent coupling constants see Refs.~\cite{Typel2005}.

On the other hand, various experiments indicate a possibility of modifications of
hadron masses and widths in the medium (e.g., see~\cite{Mettag}).
These changes might be related to a partial chiral symmetry
restoration in dense and/or hot nuclear matter, cf.~\cite{Rapp,Koch}.
Motivated by the arguments of a partial chiral symmetry restoration at high baryon densities, the Brown-Rho hadron mass scaling assumption~\cite{BrownRho}, and the equivalence theorem between different RMF schemes, Ref.~\cite{Kolomeitsev:2004ff} demonstrated how one can construct RMF models that allow to incorporate simultaneously in-medium modifications of the
baryon and meson masses and coupling constants. In this approach the coupling constants and
meson masses depend on the $\sigma$ meson field. The masses of the $\sigma$, $\omega$, and $\rho$ fields, as well as the nucleon mass, are scaled by a universal scaling function $\Phi$,
dependent on the $\sigma$ mean field. A support for the common dropping of the $N$, $\sigma$ , $\omega$, $\rho$ masses comes from the lattice QCD in the strong coupling limit~\cite{Ohnishi:2008yk}, where it was found that meson masses are approximately proportional to the equilibrium value of the
chiral condensate.  Remarkably, in the case of infinite matter the effective hadron mass ($m^{*}_h$) and the coupling constants ($g_h$) enter all relations only in the combination $m^{*\,2}_h/g_h^2$ that greatly simplifies consideration. In order to obtain  reasonable EoS, in~\cite{Kolomeitsev:2004ff} also the meson-nucleon coupling constants were  scaled with the $\sigma$ mean field. Differences in the scaling functions for the effective masses of the $\omega$  and $\rho$ fields and their coupling constants allow to get a stiffer EoS at high densities, a larger maximum mass of neutron stars and a sufficiently high critical density for the direct Urca (DU) neutrino process, as favored by ``the nuclear medium cooling scenario'' of neutron stars, cf.~\cite{Blaschke:2004vq,Grigorian:2005fn,Blaschke:2011gc,Blaschke:2013vma}.

Recently, Refs.~\cite{Dong:2009vh,Dong:2011xn} introduced the ring-diagram EOS  obtained from realistic low-momentum $NN$ interactions $V_{\rm low-k}$. Several microscopic $NN$ potentials
(CDBonn, Nijmegen, Argonne V18 and BonnA) were employed. It was demonstrated that results are largely improved with an inclusion of a Skyrme-type three-body force based on the Brown-Rho scaling, when the in-medium meson masses, particularly of $\sigma$, $\omega$ and $\rho$, are slightly decreased compared to their in-vacuum values. Best fits were done with the
Brown-Rho-Ericson (BRE) scaling, cf. \cite{Ericson}, given by $m^{*}_N/m_N =1/(1+Dn/n_0)^{1/3}$ with $D\simeq 0.35\pm 0.06$, $n$ is the baryon density, $n_0$ is the nuclear saturation density, and $m_N$ is the nucleon mass in vacuum. Coupling constants were not scaled.
Reference~\cite{Paeng:2013xya} argued on the basis of the renormalization-group analysis that while the $\rho NN$ coupling constant may fall rapidly with the density, the $\omega NN$ coupling constant should be scaled at a slower pace, being in favor of the assumption of~\cite{Kolomeitsev:2004ff} that the scalings of coupling constants are not universal. In the renormalization group approach~\cite{Paeng:2013xya} the effective nucleon mass undergoes a drop
roughly linear in the density up to a value slightly above the nuclear matter saturation density  and then stays roughly constant up to the dilaton-limit fixed point. Below we will demonstrate a similar behavior within one of our selected RMF models.

Reference~\cite{Klahn:2006ir} formulated most important constraints extracted from the empirical data, which should be fulfilled. It was shown that the models suggested
in~\cite{Kolomeitsev:2004ff} (namely, MW(nu, $z=0.65)$ and MW(nu, $z=2.9$) labeled then in~\cite{Klahn:2006ir} as KVOR and KVR models, respectively) satisfy majority of the constraints  known to that time. In the KVR model parameters were adjusted to describe the microscopic Akmal-Pandharipande-Ravenhall (APR) EoS of the Urbana-Argonne group A18 + $\delta v$ + UIX*~\cite{APR} (in relativistic HHJ-parameterization of~\cite{HHJ}) at densities below four times the
saturation density. A slightly modified parameter set (KVOR) of
this RMF model allows for a higher maximum neutron star mass. The
scaled hadron mass-couplings (SHMC) model of Refs.~\cite{Khvorostukhin:2006ih,Khvorostukhin:2008xn} is a
generalization to finite temperatures of the KVOR model used in
\cite{Kolomeitsev:2004ff,Klahn:2006ir} to describe cold nuclear matter. The SHMC model was successfully applied in Ref.~\cite{Khvorostukhin:2006ih,Khvorostukhin:2008xn,Khvorostukhin:2010aj,Khvorostukhin:2009pe}
to description of various characteristics of heavy-ion collisions
in a broad density-temperature region.

Some of the previously used constraints were recently tightened and
new constraints were formulated. At present any useful EoS of
the cold hadron matter should:
\begin{itemize}
\item[]($i$) satisfy experimental information extracted from the description of nuclear matter at low densities and  not to contradict results of microscopically based approaches,
\item[]($ii$) satisfy experimental information extracted from the description of global characteristics of atomic nuclei, reproducing baryon optical potentials at low energies, the binding energy per particle, the symmetry energy coefficient and their density derivatives at the appropriate value of the saturation nuclear density, values of the incompressibility and the effective nucleon mass,
\item[]($iii$) not to contradict the direct and elliptic flow constraints \cite{Danielewicz:2002pu} and data on the $K^+$ production in heavy-ion collisions~\cite{Lynch},
\item[]($iv$) allow for the heaviest known neutron stars PSR~J1614-2230 with the mass $1.97\pm 0.04 M_{\odot}$ and PSR~J0348+0432 with the mass $2.01\pm 0.04 M_{\odot}$
\cite{Demorest:2010bx,Antoniadis:2013pzd} ($M_\odot$ is the solar mass, $M_\odot=1.99\times 10^{33}$\,g),
\item[]($v$) avoid a too rapid cooling of the majority of the known pulsars by the direct Urca (DU) neutrino processes $n\to p+e+\bar{\nu}_e$, $p+e\to n+\nu_e$
\cite{Blaschke:2004vq,Kolomeitsev:2004ff},
\item[]($vi$) explain the gravitational mass and total baryon number of pulsar PSR J0737-3039(B) with at most 1\% deviation from the baryon number predicted for this particular object \cite{Podsiadlowski},
\item[]($vii$) yield an appropriate mass-radius relation \cite{Bogdanov:2012md,Hambaryan2014,Heinke:2014xaa}.
\end{itemize}
Additionally, the so-constrained EoS of the cold nuclear matter with an arbitrary isospin
composition should be extended to non-zero temperatures to be tested against
a wealth of data extracted from analyses of heavy-ion collisions.

Further information about the compact star mass-radius relation
can be found in Refs.~\cite{Lattimer:2012nd,Steiner:2012xt}. The
statistics of 61 measured masses of neutron stars  in binary
pulsar systems is collected in \cite{Zhang:2010qr}. In contrast to
mass determinations, there are no high-accuracy radius
measurements. In Ref.~\cite{Bogdanov:2012md} the neutron star
radius was constrained to be larger than $ 11.1$ km at the $3\sigma$ confidence level,
with the assumption that the mass of the object is $M = 1.76 M_{\odot}$, for all combinations of other
parameters. If taken in combination with the best available mass
measurement from radio timing~\cite{Verbiest}, the derived
constraints prove to be inconsistent with all EoSs except very stiff ones.
It is interesting to note that the lower bounds on the confidence intervals are still tighter.
The X-ray spin-phase-resolved spectroscopic study of three thermally emitting isolated neutron
stars and fits of highly magnetized atmospheric models~\cite{Hambaryan2014} allowed
to estimate their compactness, which would also prefer a stiff EoS. None of EoSs analyzed in
Ref.~\cite{Klahn:2006ir} is able to explain the suggested radius $R>14$~km
of the source RX~J1856.5-3754. Note that~\cite{Lattimer:2012nd}
suggested a smaller most probable radius of RX J1856.5-3754, which would be then
compatible with some known EoSs. A possible correlation between the neutron skin of
$^{208}$Pb and the neutron star radius was explored in Ref.~\cite{Erler:2012qd}.

If we denote as $M^{(n)}_{\rm DU}$ the critical mass of  the neutron star, at which the nucleon DU reactions $n\to p+e+\bar{\nu}_e$, $p+e\to n+\nu_e$ become possible,
every star with a mass only slightly heavier than $M^{(n)}_{\rm DU}$ will be
efficiently cooled by the DU-processes even in the presence of
superfluidity (except the case where a proton pairing gap is artificially increased) and becomes almost invisible for thermal detection
within a few years~\cite{Blaschke:2004vq,Grigorian:2005fn}.
The II-supernova explosion scenario~\cite{Woosley} and the population synthesis models, see \cite{Popov}, suggest that most of single neutron stars have, probably, masses below $1.5 M_{\odot}$. Therefore one may assume that majority of the pulsars, which surface
temperatures have been measured,  have masses $M\lsim  1.5 M_{\odot}$. Thus, \cite{Klahn:2006ir} suggested as a ``strong'' DU constraint that $M^{(n)}_{\rm DU}$ should be larger than $1.5 M_{\odot}$ and as a ``weak'' constraint that at least $M^{(n)}_{\rm DU} > 1.35 M_{\odot}$, where $1.35 M_{\odot}$ is taken as the mean value of the neutron star mass measured in binaries. One can not finally exclude a possibility of lower values of $M^{(n)}_{\rm DU}$ (even $M^{(n)}_{\rm
DU}<1.35 M_{\odot}$) within a more exotic explanation of the
present cooling data. However, the absence of the DU process for
typical neutron star configurations can be considered as the most
realistic scenario. Reference~\cite{Kolomeitsev:2004ff}
demonstrated that the standard RMF-based models yield very low
values for the DU threshold density and the corresponding star mass, whereas
with the help of the $\sigma$-dependent scaling of hadron masses and coupling constants the
problem can be avoided. Similar results are obtained in  some RMF
models with density dependent couplings, see~\cite{Klahn:2006ir}.

The nuclear EoSs of  various models can be characterized by
comparing the parameters in the expansion of the energy
per nucleon near the nuclear saturation density $n_0$ in terms of the
density deviation $\epsilon  = (n-n_0)/n_0$ and the asymmetry $\beta = (n_n-n_p)/n$ for small $\beta$:
\begin{align}
\mathcal{E}(n,\beta) &= \mathcal{E}_0  +\frac{K}{18}\epsilon^2
-\frac{K^{'}}{162}\epsilon^3 +...+\beta^2
\widetilde{\mathcal{E}}_{\rm sym}(n) + O(\beta^4) \,,
\label{Eexpans}
\\
\widetilde{\mathcal{E}}_{\rm sym}(n) &=\frac12\frac{\partial^2 \mathcal{E}}{\partial\beta^2}\Big|_{\beta=0}=J +\frac{L}{3}\epsilon +\frac{K_{\rm
sym}}{18}\epsilon^2...\,.
\label{Jtilde}
\end{align}
In this form the EoS is characterized at saturation by the energy per nucleon $\mathcal{E}_{0}$, the incompressibility $K$, the skewness parameter
$K^{'}$ for the isospin-symmetric part and by the symmetry energy ${J}$, the symmetry
pressure $L$, and the curvature parameter $K_{\rm sym}$ for the isospin-dependent part of $\widetilde{\mathcal{E}}_{\rm sym}(n)$.

The parameters characterizing nuclear matter properties at the nuclear saturation are only partially constrained from the data and have rather broad uncertainties, cf.~\cite{Agrawal}. One usually fixes the saturation density to be $n_0 \simeq 0.16\pm 0.015$ fm$^{-3}$ and the energy per nucleon, $ \mathcal{E}_0 \simeq -15.6\pm 0.6$\,MeV. Experimental value of the nuclear incompressibility at nuclear saturation is usually extracted from the data on the Giant
Monopole Resonance (GMR), see~\cite{Blaizot}. Reference~\cite{Shlomo}
adopted the value  $K=240\pm 20$MeV, whereas based on the most
precise up-to-date data on GMR energies of Sn and Cd isotopes,
together with a selected set of data from $^{56}$Ni to $^{208}$Pb, Ref.~\cite{Stone:2014wza} extracted a value of $K$ varied in a broad range from 250 to 315~MeV. As has been demonstrated in Ref.~\cite{Kolomeitsev:2004ff}, the maximum value of the neutron star
mass is a sharp  function of the chosen value of the (Dirac)
effective nucleon mass at the nuclear saturation $m_N^{*}(n_0)$
and a smoother function of $K$. The effective nucleon mass is
not well constrained from the data. For example, in order to describe the
spin-orbit splitting in atomic nuclei, a value below $0.64 m_N$ is
required, see~\cite{Furnstahl98}. Larger values of the effective Dirac nucleon mass are
motivated by chiral effective theory expansions
\cite{Drischler:2013iza} and by fitting the single nucleon spectra
in nuclei \cite{SapKhodel} with a large Landau mass~\footnote{In our relativistic mean field approximation the Dirac effective mass is the mass which enters the Dirac equation for fermions in the medium, whereas the Landau mass follows from equality $p_{\rm F} /m_{\rm L}^* =(d\epsilon (p)/d p)|_{p_{\rm F}}$,  where $\epsilon (p)=\sqrt{m^{*2}_{\rm D} +p^2}+V$, $V$ is the contribution from a vector potential. We will denote the nucleon Dirac effective mass simply as $m_N^*$ and the nucleon Landau mass as $m^*_{N,\rm L}$.}
$m^{*}_{N,{\rm L}} \simeq 0.9\mbox{---}1.0 m_N$. The works~\cite{Johnson} find the Landau
mass $m^{*}_{N,{\rm L}} \simeq 0.74\mbox{---}0.82 m_N$ from the analysis of
neutron scattering off lead nuclei. An extrapolation from finite
nuclei to nuclear matter of the nucleon optical potential at a low
particle energy \cite{Feldmeier} is better fitted for $m_N^{*} \sim 0.7 m_N$. The
latter values relate to the Dirac mass $m_N^{*}\simeq 0.7\mbox{---}0.8
m_N$.

Note that the symmetry energy  is often determined as
\begin{align}
\mathcal{E}_{\rm sym}(n) =\mathcal{E} (n_p=0) - \mathcal{E} (n_n=n_p)
 \,.
\label{Jn}
\end{align}
Usually the so-defined quantity $\mathcal{E}_{\rm sym}$ differs only a little from the value $\widetilde{\mathcal{E}}_{\rm sym}(n)$ introduced in Eq.~(\ref{Jtilde}).

The symmetry energy is critical for understanding  the structure of
rare isotopes, heavy-ion reactions and for many issues in
astrophysics including questions on the DU threshold, and, as we
will demonstrate below, the baryon-gravitational mass relation.
Experimental data on the density dependence of the symmetry energy
are available mostly for reactions with stable beams
\cite{LiChen}. Forthcoming experiments with more neutron-rich nuclei at
several radioactive beam facilities under construction will be able to improve
the situation. Existing constraints on the values of $L$ and
$K_{\rm sym}$ extracted in analyses of different experiments are
rather controversial.  Reference~\cite{Tsang:2012se} examined the
results of laboratory experiments that provided constraints on
the nuclear symmetry energy  for the nucleon density  $n\lsim n_0$.
Some of these constraints have been derived from the properties of
nuclei, others from the nuclear response to hadronic and
electroweak probes. The analyses of the isospin diffusion and the
$\pi^-/\pi^+$ ratio in heavy-ion collisions making use of the IBUU04
transport model constraining the symmetry energy at sub- and
supra-saturation densities are discussed in~\cite{LiChen,Li:2008fka}. Some  constraints on the density dependence of the symmetry energy ($L=106\pm 46$ MeV and
$K_{\rm sym}=127\pm 290$ MeV) were extracted from comparing
predictions for the neutron-proton elliptic-flow difference
$np$EFD and $np$EFR with the FOPI-LAND experimental data~\cite{Cozma:2013sja}. Analyzing available experimental information on neutron skins measured
for 26 stable nuclei, from $^{40}$Ca to $^{238}$U, in anti-protonic atoms,
Ref.~\cite{Centelles:2008vu} suggested a constraint that $L=55\pm
25$ MeV.
Studying the measurements of asymmetry skins in nuclei, Ref.~\cite{DanielewiczLee} arrived at constraints ${J} = (30.2-33.7)$\,MeV and $L = (35-70)$\,MeV.
The fitting of the data on the atomic masses  within the Hartree-Fock-Bogoliubov nuclear mass models, based on 16-parameter generalized Brussels' Skyrme forces~\cite{PearsonEPJA50} favours a value of $\simeq 30$\,MeV for the symmetry coefficient $J$.
The neutron skin thickness was extracted from the electric dipole response of $^{208}$Pb~\cite{Tamii}, which allowed to  better constraint the $L$--$J$ correlation suggested in~\cite{Roca-Maza}
Studying broadly different mean field interactions, Ref.~\cite{Dong:2012zza} found an empirical  relation for the symmetry energy at the saturation density, the slope parameter $L$, and the curvature parameter $K_{\rm sym}$.
Description of the symmetry energy within quantum statistical
approach  with account for the  formation of clusters allowed~\cite{Natowitz:2010ti} to join the low-density limit with quasiparticle approaches valid near the saturation density. A correlation between the symmetry energy and the neutron star radius was studied in \cite{Gandolfi:2013baa} using quantum Monte Carlo methods.
Imprint of the symmetry energy on the inner crust and strangeness content of neutron stars was analyzed in~\cite{ProvidenciaEPJA50}.

At low densities the nuclear EoS is rather well constrained from
microscopic approaches starting with realistic vacuum $NN$
potentials. One may use the variational
microscopic (APR) EoS based on the A18$+\delta
v+$UIX$^{*}$ $NN$ forces~\cite{APR} or the results of relativistic
Dirac-Brueckner-Hartree-Fock calculations based on the Bonn potentials, see~\cite{Fuchs}, Monte
Carlo simulations~\cite{Gandolfi:2013baa} or the calculations within a chiral effective field theory. Microscopic neutron matter calculations based on chiral $NN$ and $3N$ interactions were confronted with the constraints on properties of nuclear matter in~\cite{Hebeler:2013nza,Hebeler:2014ema}.

A set of constraints was also formed from a comparative analysis of
240 non-relativistic Skyrme parameterizations~\cite{Dutra}
describing nuclear matter at densities $n\lsim 3n_0$. Note  that
Skyrme functionals predict rather strong deformations for the
light Pb isotopes. According to the analysis of Ref.~\cite{Tolokonnikov:2014ija}, this contradicts to experimental data on charge radii and magnetic moments of odd Pb isotopes. Reference~\cite{Tolokonnikov:2014ija} exploited the phenomenological Fayans
functional~\cite{Fayans} with the volume part of the energy fitted
to the old Friedman-Pandharipande parameterization~\cite{FPold},
which is rather close to the parametrization of the APR EoS.
A large set of 263 RMF models was examined in~\cite{Dutra:2014qga} against several sets of constraints. Models with  field-dependent hadron
masses and coupling constants were not considered.

From experimental results on the single-particle energies of many
$\Lambda$-hypernuclei, the effective binding potential acting on $\Lambda$ in infinite nuclear matter with the strength $U_{\Lambda N}\sim -30 $MeV
was extracted in~\cite{Harada}.  The $\Sigma$-hyperon potential proves to be $U_{\Sigma N}\lsim +30$ MeV from the  observed quasi-free $\Sigma$ production spectra.  From analyses of the twin
hypernuclear formation~\cite{Aoki} and $\Xi$ production spectra~\cite{Khaustov} the potential  of $\Xi$ was estimated to be around $U_{\Xi N} \sim -15$ MeV. With an increase of the nucleon density towards the neutron star interior the nucleon chemical potential may become as large as the
effective hyperon mass. If so, new Fermi seas of the hyperons should begin to fill in~\cite{Ambarzumyan,Bethe,Moszkowski}. As the result, the EoS
becomes softer that causes a decrease of  the maximum mass of the
neutron star with hyperons. The latter may come into  contradiction with
observations of the most massive neutron stars. This is one part
of the ''hyperon puzzle'', see~\cite{SchaffnerBielich:2008kb,Djapo:2008au}. The puzzle was
extensively investigated within the RMF models, see in~\cite{Glendenning,Sedrakian07}. Most of the works predicted masses $\lsim 1.6 M_{\odot}$ for non-rotating stars~\cite{Bhowmick:2014pma,Massot:2012pf}. Masses $\lsim 1.8
M_{\odot}$ were obtained in non-relativistic phenomenological
models~\cite{Balberg}. Fully microscopic models based on
hyperon-nucleon potentials, which include the repulsive three-body
forces, also predict maximum masses for the neutron stars below
the measured values~\cite{Schulze,Logoteta}. A repulsive
three-body hyperon-nucleon force is needed to reproduce the ground
state properties of medium-light $\Lambda$ hypernuclei~\cite{Lonardoni}. Reference~\cite{Weissenborn:2011ut} demonstrated
that the combined repulsive interactions mediated by the $\omega$
and $\phi$ mesons cause the hyperons to appear only at high densities. The chiral quark-meson coupling model~\cite{Katayama:2012ge} exploited that the baryon structure
variation in matter is reflected in the $\sigma$-field dependence
of the $g_{\sigma B}^{*}$ effective coupling parameterized   in
the linear form~\cite{Miyatsu:2013hea}. The resulting decrease of
$g_{\sigma B}^{*}$ effective coupling with the density increase
allowed to describe neutron stars with masses up to $1.95
M_{\odot}$. Reference~\cite{Colucci:2014wda} used the RMF model
with a density-dependent parametrization of the nucleon-meson
couplings of Ref.~\cite{Lalazisis} to constrain density dependence
of the hyperon-nucleon coupling constants.   To satisfy the requirement
that the lower bound on the maximum mass of a compact star is  $2
M_{\odot}$ Ref.~\cite{vanDalen:2014mqa} derived constraints on the
$\sigma\Lambda$ and $\sigma\Sigma$ coupling constants. References
\cite{Bednarek:2009at,Bednarek:2011gd,Gusakov:2014ota} used a
non-linear model involving two additional hidden strangeness
scalar $\sigma^*$ and vector $\phi$ mesons coupled to hyperons
and quartic terms involving vector meson fields. At this price the
authors  fulfill the constraint on the maximum neutron star mass
$M_{\rm max}> 2 M_{\odot}$.

Another part of the ``hyperon puzzle'' is that the critical
densities, at which the first hyperon species occurs in the
neutron star matter, are low in all models ($\sim 3 n_0$), if
one uses coupling constants constrained by the SU(6) symmetry, see~\cite{Weissenborn:2011kb,KV03},
and when the hyperons arise,
efficient DU reactions on hyperons, e.g. $\Lambda\to
p+e+\bar{\nu}_e$, $p+e\to \Lambda +\nu_e$, are switched on. Thus,
there may appear a problem with fulfillment of the ``strong'' DU
constraint. The problem might be resolved, if one exploits large hyperon pairing gaps or
imposes more
general SU(3) symmetry relations instead of SU(6) ones
\cite{Weissenborn:2011ut,Lopes:2013cpa}, since then with the help
of an extra fitting parameter one may decrease values of
attractive coupling constants. Below we will suggest another solution of
the problem performed  within SU(6) symmetry approach.

In this paper our main goal is to construct some effective models
of the EoS of the cold baryon matter that would incorporate
the decrease of hadron masses and the change of coupling constants with an increase of the
baryon density and, simultaneously, would fulfill various
constraints known from analyses of atomic nuclei, heavy-ion
collisions and neutron stars. The work is organized as follows. In
Sect.~\ref{GenForm} we formulate a general formalism describing
our RMF models allowing for scaling of the field mass-terms and
coupling constants with $\sigma$ field variable $f\propto
\sigma\chi_{\sigma N}(\sigma)$, where $\chi_{\sigma N}(\sigma)$ is
the scaling of the $\sigma N$ coupling constant. In Sect.~\ref{Setup} we
formulate several specific models without an inclusion of the
strange content.  More specifically we consider two types of
modifications of the previously studied KVOR model. In one type of
models (KVORcut) the EoS stiffens provided the function $f(n)$ stops growing above some chosen density larger than the saturation density.
Other model (MKVOR) demonstrates the stiffening of the EoS in the $\beta$-equilibrated neutron star matter while keeping it rather soft in the isospin symmetric matter (ISM). This is achieved, on one hand, by the proper choice of the scaling functions for $\omega$ and $\sigma$ meson contributions  and, on the other hand, by the quenching of the scalar field growth in the asymmetric nuclear matter induced by the scaling of the $\rho$ meson contribution.
Some results of the MKVOR model were briefly reported in Ref.~\cite{MKVshort}. In Sect.~\ref{Thermochar} we describe
some equilibrium characteristics of our EoSs. In Sect.~\ref{ExpConstraints} we demonstrate how our EoSs fulfill various
experimental constraints. Then in Sect.~\ref{Hyperons} we account for a possible appearance
of hyperons. Interaction with $\phi$-meson mean
field is included. In Conclusion we formulate our results.
Matching of the RMF EoSs and the crust EoS is considered in Appendix~\ref{app:crust}.
Cumbersome expressions for Landau parameters are deferred to
Appendix~\ref{app:Landpar}. Throughout the paper we use units $\hbar =c=1$.

\section{General formalism}\label{GenForm}

We start with a generalized RMF model, where effective hadron
masses and coupling constants are assumed to be $\sigma$
field-dependent \cite{Kolomeitsev:2004ff}. The Lagrangian density
is given by the sum of baryon $\mathcal{L}_{\rm bar}$, meson mean-field  $\mathcal{L}_{\rm mes}$, and lepton $\mathcal{L}_l$ terms:
\begin{eqnarray}
\mathcal{L} = \mathcal{L}_{\rm b} + \mathcal{L}_{\rm mes} + \mathcal{L}_l\,.
\label{lagKV}
\end{eqnarray}
We include the full ground-state octet of baryons $b$ with nucleons $N=(p,n)$ (proton
$p$ and neutron $n$) and  hyperons $H=(\Lambda^0, \Sigma^{\pm,0}, \Xi^{-,0})$ with the parameters listed in Table~\ref{tab:particles},
\begin{eqnarray}
&&\mathcal{L}_{\rm b} = \sum_b \bar{\Psi}_b\big[i D_\mu^{(b)} \gamma^\mu - m_b \Phi_b \big] \Psi_b, \quad b = (N, \Lambda, \Sigma^{\pm,0}, \Xi^{-,0})\,,
 \nonumber \\
&&D_\mu^{(b)} = \partial_\mu + i g_{\om b} \chi_{\om b} \omega_\mu + i g_{\rho b} \chi_{\rho b}\vec{t} \vec\rho_{\mu} + i g_{\phi b} \chi_{\phi b} \phi_\mu \,.
\label{Lag-bar}
\end{eqnarray}
Here $\Psi_b$ are the baryon bispinor, $\gamma^\mu$ denote the Dirac $\gamma$-matrices and $\vec{t}$ is the baryon isospin operator. Greek indices run over $0,1,2,3$, Latin indices
run over $1,2,3$. Within the approach of Ref.~\cite{Kolomeitsev:2004ff} the
coupling constants $g_{\sigma b}$, $g_{\omega b}, g_{\rho b}, g_{\phi b}$ are made dependent of the $\sigma$ field  with the help of scaling functions as $g_{\sigma b} \chi_{\sigma b}$, $g_{\omega b} \chi_{\omega b}$, $g_{\rho b} \chi_{\rho b}$, $ g_{\phi b}
\chi_{\phi b}$.
Among the meson mean fields we include the scalar meson and vector mesons, $m=(\sigma, \omega, \rho,\phi)$, where the $\phi$ meson is operative, if hyperons are incorporated into the scheme,
\begin{align}
\mathcal{L}_{\rm mes}  &= \half \partial_\mu \sigma \partial^\mu \sigma
- \half m_\sigma^{2}\Phi_\sigma^2 \sigma^2  - {U}(\sigma)
-\quart\om_{\mu \nu} \om^{\mu \nu} + \half m_\omega^{2}\Phi_\omega^2 \om_\mu \om^\mu
\nonumber \\
& - \quart\rho_{\mu \nu} \rho^{\mu \nu}
+ \half m_\rho^{2}\Phi_\rho^2 \vec{\rho}_\mu \vec{\rho}^{\,\mu}
- \quart\phi_{\mu \nu} \phi^{\mu \nu}+ \half m_\phi^{2}\Phi_\phi^2 \phi_\mu \phi^\mu
,
\nonumber \\
&\om_{\mu \nu} = \partial_\nu \om_\mu - \partial_\mu \om_\nu, \quad \vec{\rho}_{\mu
    \nu} = \partial_\nu \vec{\rho}_\mu -  \partial_\mu \vec{\rho}_\nu\,,\quad
 \phi_{\mu \nu} = \partial_\nu \phi_\mu - \partial_\mu \phi_\nu \, .
\label{Lag-mes}
\end{align}
By omitting, for simplicity, the $\rho\rho$ interaction term, we disregard a possibility of the charged $\rho$ meson condensation considered in~\cite{Kolomeitsev:2004ff}. Here $U(\sigma)$ is a self-interaction term of the $\sigma$ field. The bare masses of baryons, $m_b$, and mesons, $m_{m}$, are replaced in Eqs.~(\ref{Lag-bar}) and (\ref{Lag-mes}) by the effective masses $m_b^*=m_b\Phi_b$, $m_{m}^*=m_m\Phi_m$.
The Lagrangian
\begin{align}
\mathcal{L}_l = \sum_l \bar{\psi}_l (i \partial_\mu \gamma^\mu - m_l) \psi_l
\end{align}
contains contributions of the light leptons, electrons and muons, $l=e,\mu$,  and $\psi_l$
stands here for the lepton bispinor. The lepton masses $m_l$ are the bare
masses, see Table~\ref{tab:particles}.

\begin{table}
\caption{Particle properties.}
\centering
\begin{tabular}{ccccccccccc}
\hline \hline
& n & p & $\Lambda^0$ & $\Sigma^-$ & $\Sigma^0$ & $\Sigma^+$ & $\Xi^-$ & $\Xi^0$ & $e^-$ & $\mu^-$ \\ \hline 
$m [{\rm MeV}]$ & 938 & 938 & 1116 & 1193  & 1193 & 1193 & 1318 & 1318 & 0.5 & 105
\\ \hline \rule[-2ex]{0pt}{5.5ex}
$t_{3}$  & $-\frac{1}{2}$ & $\frac{1}{2}$ & 0 & -1 & 0 & 1 & $-\frac{1}{2}$&$\frac{1}{2}$ & -- &--\\
\hline\hline
\end{tabular}
\label{tab:particles}
\end{table}
We will assume that the $\sigma$ field dependence enters the scaling function $\chi_{mb}$ and $\Phi_{b(m)}$ through an auxiliary variable
\begin{align}
f = g_{\sigma N} \chi_{\sigma N} (\sigma) \frac{\sigma}{m_N}\,.
\label{f-def}
\end{align}
Following Ref.~\cite{Kolomeitsev:2004ff} we  exploit  the universal scaling for the field-mass functions
\begin{align}
\Phi_N (f)=\Phi_m (f)=1-f.
\label{PhiN}
\end{align}
We suppose that
\begin{align}
\chi_{\omega H}(f) = \chi_{\omega N}(f)\,,\,\,\, \chi_{\rho H}(f)
= \chi_{\rho N}(f)\,,
\end{align}
and express the hyperon-mass scaling function $\Phi_H$ through the nucleon-mass scaling function as
\begin{align}
\Phi_H (f)=\Phi_N\big(g_{\sigma H}\chi_{\sigma H}\frac{\sigma}{m_H}\big)\equiv
 \Phi_N\big(x_{\sigma H}\xi_{\sigma H}\, \frac{m_N}{m_H}\,f\big)\,,
 \quad \xi_{\sigma H}=\frac{\chi_{\sigma H}}{\chi_{\sigma N}}\,,
\end{align}
where $\xi_{\sigma H}$ is a function of $f$.

Taking into account the equations of motion for vector fields, the
energy density of the cold infinite matter with an arbitrary
particle  composition is recovered from the Lagrangian of the model
in the standard way, see Ref.~\cite{Kolomeitsev:2004ff}:
\begin{align}
E[f,\{n_i\}] &=
\sum_b E_{\rm kin}( m_b \Phi_b (f),p_{{\rm F},b })
+  \sum_{l = e, \mu} E_{\rm kin}( m_l,p_{{\rm F},l })
\nonumber\\
& +\frac{m_N^4 f^2}{2 C_\sigma^2 } \eta_\sigma(f)  +
\frac{C_\om^2}{2 m_N^2 \eta_\om(f)} \big(\sum_b x_{\om b} n_b\big)^2 +
\nonumber\\
&+ \frac{C_\rho^2}{2 m_N^2 \eta_\rho(f)} \big(\sum_b x_{\rho b} t_{3 b} n_b \big)^2
+  \frac{C_\phi^2}{2 m_N^2 \eta_\phi(f)} \big( \sum_{H} x_{\phi H} n_H \big)^2 \,,
\nonumber\\
E_{\rm kin}(m,p_\rmF)& =\frac{1}{8 \pi ^2}
\Big(p_\rmF \sqrt{m^2 + p_\rmF^2} (m^2+2 p_\rmF^2)-m^4 {\rm arcsinh}(p_\rmF/m)  \Big)\,,
\label{Efunct}
\end{align}
where the Fermi momentum of species $i$ is $p_{\rmF,i}=(3\pi^2\,n_i)^{1/3}$
and we introduce the dimensionless coupling constants $C_M = g_{MN} m_N/m_M$, $M=(\sigma,\om,\rho)$, and the ratios of coupling constants $x_{M H} = {g_{M H}}/{g_{M N}}$.
Since $g_{\phi N}=0$ we have to define the ratios $x_{\phi B}$ through $g_{\om N}$: $x_{\phi B} = g_{\phi B}/g_{\om N}$, the $\phi$-field contribution to the energy density is determined by the constant $C_\phi=C_\om m_\om/m_\phi$. Here we take $m_\om = 783\,{\rm MeV}$, $m_\phi
= 1020\,{\rm MeV}$. Masses of other mesons and their
coupling constants enter the energy density only via the combinations $C_M$.
The scaling functions $\Phi_\om$, $\chi_\om$ and $\Phi_\rho$, $\chi_\rho$ for vector meson fields enter only in the combinations
\begin{eqnarray}
\eta_\om (f) = \frac{\Phi_\om^2 (f)}{\chi_{\om N}^2 (f)}\,, \quad \eta_\rho (f) = \frac{\Phi_\rho^2 (f)}{\chi_{\rho N}^2 (f)}\,.
\end{eqnarray}
Therefore, we actually do not need to determine $\Phi_\om$, $\chi_\om$ and $\Phi_\rho$, $\chi_\rho$ separately, but only $\eta (f)$ combinations. The potential ${U}(\sigma)$ of the $\sigma$ self-interaction might be hidden in the definition of $\eta_{\sigma}(f)$ as
\begin{eqnarray}
\eta_{\sigma}(f)=\frac{\Phi_{\sigma}^2[\sigma(f)]}{\chi_{\sigma N}^2[\sigma(f)]} + \frac{ 2 \, C_{\sigma}^2}{m_N^4 f^2}  {U}[\sigma(f)]\,.
\end{eqnarray}
The equation of motion for the remaining field variable $f$ follows from the minimization of the energy density $\frac{\partial E[f, \{n_i\}]}{\partial f}=0$,
which in the explicit form reads
\begin{eqnarray}
&&\frac{m_N^4 f}{C_\sigma^2 } \eta_\sigma(f)
=-\sum_{b}m_b\Phi'_b(f)\rho_{\rm S}(m_b\Phi_b(f),p_{\rmF,b})-m_N\rho_\eta(f,\{n_i\})\,.
\label{eq_fn}
\end{eqnarray}
Here the first term on the r.h.s is the standard source of the scalar field associated with  the baryon scalar density $\rho_{\rm S}$:
\begin{eqnarray}
\rho_{\rm S}(m,p_\rmF) = \frac{1}{2 \pi ^2} \big(m p_\rmF \sqrt{m^2 + p_\rmF^2}
- m^3 {\rm arcsinh}(p_\rmF/m)
\big)   \,.
\label{rhoS}
\end{eqnarray}
The second term  arises due to the scaling functions $\eta_i$,
\begin{eqnarray}
\rho_{\eta}(f,\{n_i\}) &=& \frac{m_N^3 f^2}{2 C_\sigma^2 } \eta_\sigma'(f)
 - \frac{C_\om^2 \eta'_\om(f)}{2 m_N^3 \eta^2_\om(f)} \big(\sum_b x_{\om b} n_b\big)^2  +
\nonumber\\
&-& \frac{C_\rho^2 \eta'_\rho(f)}{2 m_N^3 \eta^2_\rho(f)} \big(\sum_b x_{\rho b} t_{3 b} n_b \big)^2
- \frac{C_\phi^2 \eta'_\phi(f)}{2 m_N^3 \eta^2_\phi(f)} \big( \sum_{H} x_{\phi H} n_H \big)^2 .
\label{rho-eta}
\end{eqnarray}
Solving Eq.~(\ref{eq_fn}) we determine the mean scalar field $\bar{f}$ as a function of baryon densities.

The coupling constants of hyperons to vector mesons can be related to those of nucleons with the help of the SU(6) symmetry relations~\cite{Weissenborn:2011kb}:
\begin{align}
&g_{\om \Lambda} = g_{\om \Sigma}= 2 g_{\om \Xi}=\frac{2}{3} g_{\om N} \,,
\quad
g_{\rho\Lambda}=0\,, \quad  g_{\rho \Sigma} = 2g_{\rho \Xi} = 2g_{\rho N}\,, \nonumber\\
&2 g_{\phi \Lambda} = 2 g_{\phi \Sigma} = g_{\phi \Xi} = -\frac{2 \sqrt{2}}{3} g_{\omega N}\,,
\quad g_{\phi N} = 0.
\label{gHm}
\end{align}
One could, of course, use more general relations dictated by the SU(3) symmetry~\cite{Weissenborn:2011kb}, which contain one additional parameter, but we prefer a more traditional approach here. The extension is though straightforward.

In our approach the scalar meson coupling constants are constrained by hyperon binding energies in the ISM at saturation, which are deduced from extrapolation of hypernucleus data.
For the given hyperon binding energy per nucleon $\mathcal{E}_{\rm bind}^{H}$,  the $\sigma H$ and $\om H$ coupling constants must be correlated in our model as
\begin{eqnarray}
\mathcal{E}_{\rm bind}^{H}(n_0)=\frac{C_{\omega}^2}{m_N^2} x_{\omega H} n_0
-x_{\sigma H}\,\xi_{\sigma H}(\bar{f}_0)\, [m_N-m_N^{*} (n_0)]\,,
\label{EHbind}
\end{eqnarray}
where  $\bar{f}_0$ is the solution of Eq.~(\ref{eq_fn}) in the ISM at
saturation, $n_p=n_n=n_0/2$.
We use the results of Refs.~\cite{Hashimoto06,Dabrowski99,Khaustov}
$$
\mathcal{E}_{\rm bind}^{\Lambda }(n_0) = -28 \,{\rm MeV},\quad
\mathcal{E}_{\rm bind}^{\Sigma }(n_0) = 30 \,{\rm MeV},\quad
\mathcal{E}_{\rm bind}^{\Xi }(n_0) = -15 \,{\rm MeV}\,.$$

In our model the composition of the neutron star core is determined by conditions of the $\beta$-equilibrium imposing the relation among the particle chemical potentials
\begin{align}
\mu_i=\mu_n - Q_i\,\mu_e \,,\quad
\mu_i=\frac{\partial}{\partial n_i} E[\bar{f},\{n_i\}]\,,\quad i=b\,,
\label{chempot-i}
\end{align}
and by the electro-neutrality condition
\begin{align}
\sum_{i} Q_i\,n_i=0\,,
\label{electroneut}
\end{align}
where  $Q_i$ is the charge of the particle of type $i$.
Solving Eqs.~(\ref{chempot-i}) and (\ref{electroneut}) together one can express the particle densities $n_i$ through the total baryon density $n = \sum_{b} n_b$.

Once all particle densities and the chemical potentials are known the pressure of the
matter in the $\beta$-equilibrium (BEM) can be calculated from the expression
\begin{align}
P[n]=\sum_i\mu_i\, n_i -E[\bar{f}(n),\{n_i\}]\,.
\label{press}
\end{align}

For densities $n< 0.7 n_0$ we match the RMF EoS with the crust EoS, as described in Appendix~\ref{app:crust}. The final neutron star configuration follows from the solution of the
Tolman--Oppenheimer--Volkoff equation.

\section{Setup of the RMF models without hyperons}\label{Setup}

In this section we define the scaling functions and input parameters for several RMF models. First we demonstrate how our generalized energy density functional (\ref{Efunct}) reproduces results of the traditional Walecka (W) model, non-linear Walecka (NLW) models and the KVOR model developed in Ref.~\cite{Kolomeitsev:2004ff}. Then we propose the method how one can make the EoSs in a certain RMF model stiffer, restricting a decrease of the nucleon mass with a density increase. The mechanism is a purely phenomenological approach in order to obtain models which comply with the constraints known from  the experimental data. Finally we formulate our novel RMF model, labeled as MKVOR, which properties we will discuss in details in the subsequent sections.

\subsection{Traditional models and the KVOR model}

The standard W model~\cite{Walecka1974} with inclusion of $\rho$ mesons  can be recovered from Eq.~(\ref{Efunct}) setting
\begin{eqnarray}
\eta_\sigma(f) = \eta_\om(f) = \eta_\rho(f) = 1, \quad U(f) = 0, \quad \Phi_N(f) = 1-f\,,
\label{W-scale}
\end{eqnarray}$\Phi_M=1$\, for the meson fields.
The hyperon and $\phi$-meson contributions are suppressed.
The W model can reproduce the saturation phenomenon in the ISM with the appropriate choice
of $\mathcal{E}_0$ and $n_0$, but only at cost of a too high value of the incompressibility and a too small value of the nucleon effective mass, see Table~\ref{tab:sat-param}. The values of the corresponding coupling constants are listed in Table~\ref{tab:param-1}, cf.~\cite{Matsui}.

\begin{table}
\setlength{\arraycolsep}{1mm}
\caption{Characteristics of the W, NLW and KVOR  EoS at saturation.}
\begin{center}
\begin{tabular}{lcccccccc}
\hline\hline
\raisebox{-.37cm}[0cm][0cm]{EoS}& $\mathcal{E}_0$ & $n_0$ & $K$ & $m_N^*(n_0)$ & ${J}$ & $L$ &$K'$ & $K_{\rm sym}$
\\ \cline{2-9}
&           [MeV] & [fm$^{-3}$] & [MeV] & $[m_N]$  & [MeV] & [MeV] &[MeV] & [MeV]
\\ \hline
W  & $-15.76$ & 0.193 & 546.59 &  0.56 & 33.67 & 109 & -1880& 74\\
NLW& $-16$ & 0.15 & 210 &  0.85 &  29.51 & 78 &  632 & -20\\
KVOR & $- 16$ & 0.16 & 275 &  0.805 &  32 & 71 &  422& -85\\
\hline\hline
\end{tabular}
\end{center}
\label{tab:sat-param}
\end{table}

The mentioned deficits of the W model can be cured by the inclusion of two extra terms of a
$\sigma$ field self-interaction~\cite{Boguta77}
\begin{eqnarray}
\eta_\om(f) = \eta_\rho(f) = 1,  \quad \Phi_N(f) = 1-f,\quad U(f) = m_N^4 (\frac{b}{3} f^3 + \frac{ c}{4} f^4)\,,
\label{NLW-scale}
\end{eqnarray}
 $\Phi_M=\chi_{MN}=1$\, for the meson fields.
Hyperon and $\phi$-meson contributions are suppressed. We will denote this model as the non-linear Walecka (NLW) model.
Two extra parameters $b$ and $c$ allow to choose appropriate values
of the incompressibility and the effective nucleon mass at $n_0$.
Below we use the parameter set for model 2 from Ref.~\cite{Cubero:1987pr}, see Table~\ref{tab:param-1}. Note that they use $m_N=939$~MeV there. The corresponding saturation properties of this EoS listed in Table~\ref{tab:sat-param} are quite reasonable.
With an account for additional soft-pion modes this model appropriately describes heavy-ion collision data at collision energies $\lsim 2\,$GeV$/A$, see~\cite{Voskresensky:1993ud}.

\begin{table}
\caption{Parameters of the W, NLW and KVOR  models.}
\begin{center}
\begin{tabular}{lcccccc}
\hline\hline
EoS& $C_\sigma^2$ & $C_\om^2$ & $C_\rho^2$ & $b\cdot 10^3$ & $c\cdot 10^4$ & $z$ \\
\hline
W    & 266.90 & 195.70 & 54.710 & 0        &  0     & --\\
NLW  & 183.68 & 64.545 & 100.00 & 17.788   & 396.74 & --\\
KVOR & 179.56 & 87.600 & 100.64 & 7.7346   & 3.4462& 0.65 \\
\hline\hline
\end{tabular}
\end{center}
\label{tab:param-1}
\end{table}


Reference~\cite{Kolomeitsev:2004ff} constructed the RMF model with scaled hadron masses and coupling constants such that it matches  the APR EoS (in the  relativistic HHJ parameterization of Ref.~\cite{HHJ}) up to $n\lsim 4n_0$. To fulfill the DU constraint Ref.~\cite{Kolomeitsev:2004ff} introduced the scaling function $\eta_\rho(f)\neq 1 $. Thereby, the  MW(nu, $z=2.9$) model was constructed, see Eq.~(63) in~\cite{Kolomeitsev:2004ff}, being labeled as the KVR model in Ref.~\cite{Klahn:2006ir}. The idea behind KVOR modification of the KVR model was to demonstrate that introducing the additional scaling function $\eta_{\om}\neq 1$ one can increase the maximum value of the neutron star mass not changing significantly the KVR EoS for $n\lsim 4n_0$. In Ref.~\cite{Kolomeitsev:2004ff} the model was labeled as MW(nu, $z=0.65$), see Eq. (58) there. We will further exploit notation KVOR of Ref.~\cite{Klahn:2006ir}. The scaling functions of the KVOR  model are as follows:
\begin{eqnarray}
\eta_\om(f) = \frac{1 + z \bar{f}_0}{1 + z f}\,,\quad
\eta_\rho (f)={\eta_\om (f)}
\left[{\eta_\om (f) + 4\,\frac{C_\om^2}{C_\rho^2}\,(\eta_\om (f)-1)}\right]^{-1}\,.
\label{KVOR_etar}
\end{eqnarray}
The saturation properties for this model are given in Table~\ref{tab:sat-param}.
For the chosen parameters of the model, cf.~Table~\ref{tab:param-1}, the scaling function
$\eta_\rho$ has a pole at $f\simeq 0.7$, which is harmless at vanishing temperatures, since $f(n)$ remains smaller than 0.7. However it was shown in Ref.~\cite{Khvorostukhin:2006ih} that at
high temperatures $\eta_\rho$ may become singular. To cure this problem Ref.~\cite{Khvorostukhin:2006ih} suggested to exploit the Taylor expansion of the function $\eta_\rho$ in terms of $1-\eta_{\om}$, keeping first ten terms. For $T=0$ this expansion
fully reproduces the KVOR result.
In the scalar meson sector we use in Eq.~(\ref{Efunct})
\begin{eqnarray}
\eta_\sigma = 1 + 2 \frac{C_\sigma^2}{ f^{2}}\,  \big(\frac{b}{3} f^3 +
\frac{c}{4} f^4\big)\,,\quad \chi_{\sigma N}=\Phi_{\sigma}\,,
\label{KVOR_etas}
\end{eqnarray}
which is equivalent to the choice used in Ref.~\cite{Kolomeitsev:2004ff}.
Hyperons and $\phi$-meson contributions are disregarded.

\subsection{$\omega$-scaling cut models}

The modern experimental data indicate that the value of the
maximum neutron star mass should be sufficiently high. It invites us to
search for a simple way to stiffen the EoS at large densities without altering
it for densities less than several $n_0$.
In Ref.~\cite{MKV15-cut} we demonstrated that within an RMF model the EoS becomes stiffer for $n>n^*$ if
a growth of the scalar field as a function of the density is quenched  and the nucleon mass becomes weakly dependent on the density for $n>n^*$. In~\cite{MKV15-cut} such a quenching was achieved by adding to the scalar potential a function of $f$ rapidly rising at $f\sim f^*$, where $f^*$ is determined by $n^*$.
We demonstrate now that the same effect can be reached by the appropriate modification of the $\eta_\om(f)$ function. We will show that the steeper is the
change of $\eta_{\om}(f)$ at $f\sim f_\om$ and the smaller value $f_\om$ is taken,
the larger becomes the maximum neutron star mass computed within
the given model.

\subsubsection{Wcut model}

\begin{table}
    \caption{Parameters of the KVOR and three KVORcut models with the $\omega$ scaling cut.}
    \begin{center}
        \begin{tabular}{cccccccccc}
            \hline\hline
EoS & $C_\sigma^2$ & $C_\omega^2$ & $C_\rho^2$ & $b\cdot 10^3$ & $c \cdot 10^4$ & $f_\om$ & $a_\omega$ & $b_\omega$\\
\hline
Wcut     & 266.90 & 195.70  & 54.710 & 0        & 0      & 0.56 & $-0.5$ & 53.30 \\
\hline
NLWcut04 & 183.68 & 64.545  & 100.00 & 17.788   & 396.74 & 0.4  & $-0.5$ & 50.00 \\
NLWcut03 & 183.68 & 64.545  & 100.00 & 17.788   & 396.74 & 0.3  & $-0.5$ & 50.00 \\
NLWcut02 & 183.68 & 64.545  & 100.00 & 17.788   & 396.74 & 0.25 &$-0.5$  & 50.00 \\
\hline
KVORcut04 & 179.56 & 87.600 & 100.64 & 7.7346 & 3.4462 & 0.454 & $-$0.5 & 55.76 \\
KVORcut03 & 179.56 & 87.600 & 100.64 & 7.7354 & 3.4161 & 0.365 & $-$0.5 & 46.78\\
KVORcut02 & 184.26 & 87.594 & 100.64 & 9.9934 & -76.383& 0.249 & $-$0.2 & 74.55 \\
\hline\hline
        \end{tabular}
    \end{center}
    \label{tab:cut-model}
\end{table}

Let us start with the simplest W model and introduce  the scaling function
\begin{eqnarray}
\eta_\om(f) = 1 + \frac{a_\om}{2} [1 + \tanh(b_\om(f - f_\om))],
\label{Wal_scaling}
\end{eqnarray}
which jumps within a narrow interval $| f-f_\om |\sim 1/b_\om\ll 1 $  from unity to the value $\eta_{\om}\simeq 1+a_{\om}$.
The corresponding parameters are listed in Table~\ref{tab:cut-model}.
As in the case of the W model, we put $\Phi_M =\chi_{\sigma N}=\chi_{\om N}=\chi_{\rho N}=1$.  At the nuclear saturation density the Wcut model produces the same saturation parameters as the original
W model. Hyperons and $\phi$-meson contributions are excluded.
We will denote this modification of the W model  as the Wcut model.

\begin{figure}
\centering
\includegraphics[width = 11cm]{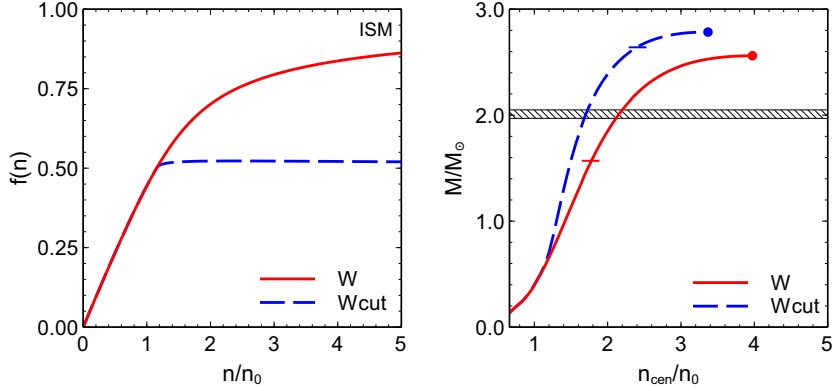}
\caption{
Left panel: The scalar field $f$ as a function of the nucleon density in the ISM
for the W model with parameters in Table~\ref{tab:param-1} and for the Wcut model     defined by Eq.~(\ref{Wal_scaling}).
Right panel: Neutron star mass versus the central density for the W and Wcut models.
Dashes mark the threshold star mass and density for the beginning of the DU reactions. Bold dots indicate the maximum neutron star masses.
The horizontal band shows the uncertainty range for the mass of PSR~J0348+0432 ($2.01\pm 0.04
M_{\odot}$).}
\label{fig:Wal}
\end{figure}

The scalar field $f$ as a function of the nucleon density in the ISM is shown in Fig.~\ref{fig:Wal}\,(left panel) for the W and Wcut models. As we see from this figure,
the $f(n)$ dependence changes sharply in the Wcut model for $n> 1.1 n_0 $ and $f(n)$ saturates at the value $f\simeq 0.52$. The neutron star masses as functions of the neutron star central density for these two models are shown on the right panel in Fig.~\ref{fig:Wal}. The value of the maximum mass for the Wcut model, $M_{\max}=2.78 M_\odot$, proves to be higher than that for the W model, $M_{\max}=2.56 M_\odot$. The DU threshold density and mass, $n^{(n)}_{\rm DU}=2.42$ and $M^{(n)}_{\rm DU}=2.64M_{\odot}$, are also increased for the Wcut model compared to the corresponding values, $n^{(n)}_{\rm DU}=1.78$, $M^{(n)}_{\rm DU}=1.57 M_{\odot}$, for the original W model.

\subsubsection{NLWcut model}

Now  we apply the same scaling (\ref{Wal_scaling}) to the NLW model. We consider three modifications NLWcut04, NLWcut03, and NLWcut02 with the parameters listed in Table~\ref{tab:cut-model}.

\begin{figure}
\centering
\includegraphics[width = 11cm]{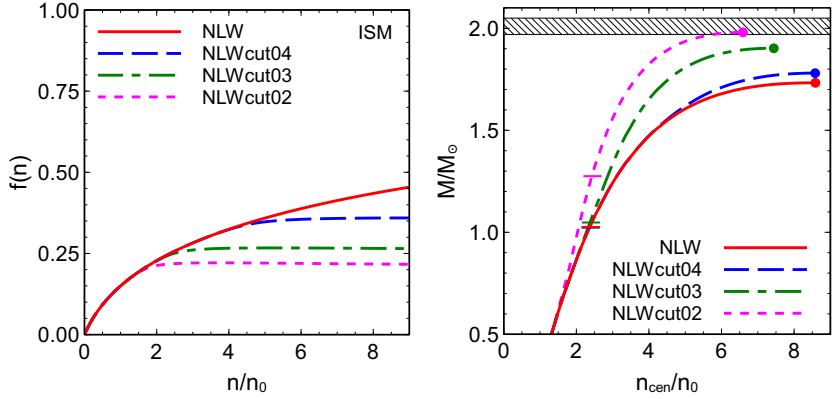}
\caption{Left panel: Scalar field $f$  as a function of the nucleon density in the ISM for the NLW and several NLWcut models with parameters given in Table~\ref{tab:cut-model}.
Right panel: Neutron star masses versus the central density for the NLW and NLWcut models.
Dashes denote the thresholds of the DU reactions. Notations of the models are given in the plot legend.
    }\label{NWal_f}
\end{figure}

The scalar field $f$  as a function of the density is shown in Fig.~\ref{NWal_f}\,(left) for the original NLW model and the NLWcut models. As we see, the $f(n)$ dependence changes sharply for the NLWcut04, NLWcut03 and NLWcut02 models for $n \gsim 3.8 n_0, 2 n_0, 1.4 n_0 $, respectively, and at higher densities $f(n)$ remains constant at values 0.36, 0.27, and 0.22.
The maximum neutron star masses for these models are shown as functions of the central density
on the right panel in Fig.~\ref{NWal_f}.
As in the case of the Wcut model, the values of the maximum masses for the NLWcut models prove to be higher than that for the original NLW model ($M_{\rm max}=1.73\,M_\odot$). The maximum mass increases with a decrease of $f_{\om}$. The curve $M(n_{\rm cen})$ for the NLWcut02 model enters the band $2.01\pm 0.04\, M_{\odot}$, corresponding to the pulsar PSR~J0348+0432. The masses corresponding to the DU thresholds also increase  with a decrease of $f_{\om}$ but remain too low to satisfy the even weaker DU constraint, $M^{(n)}_{\rm DU}>1.35\, M_{\odot}$.

We discuss the Wcut and NLWcut models only to demonstrate how with the help of the simple $\om$-cut procedure one is able to increase easily the maximum neutron star mass and the threshold DU density. However, the models fail to fulfill some other constraints. For example, the Wcut model does not pass the constraint from the particle flow in heavy-ion collisions. All Wcut and NWLcut models do not pass the constraint on the baryon mass vs. the gravitational mass of a neutron star. Therefore now we will try to apply the cut procedure to construct a more realistic RMF model.

\subsubsection{KVORcut and MKVOR models}

We construct now a set of new KVORcut models which reproduce the EoS of the KVOR model up to a certain value of the density and yield stiffer EoSs at higher densities. To obtain such a behavior we modify only the scaling function of the $\omega$ meson for $f > f_\omega$
as follows
\begin{eqnarray}
\eta_\omega(f) = \left(\frac{1 + z \bar{f}_0}{1 + z f}\right) +
\frac{a_\omega}{2} \big[1 + \tanh(b_\om (f - f_\om))\big]\,.
\label{KVORcut_etao}
\end{eqnarray}
We consider three choices for $f_{\om}$: models KVORcut02, KVORcut03, and KVORcut04.
The parameters of the models are listed in  Table~\ref{tab:cut-model}. Small  changes in the parameters appear only, if $f_{\om}$ is taken too close to $\bar{f}_0=f(n_0)$. We keep $z = 0.65$ for all KVORcut models.

The input characteristics of the EoS at saturation ($\mathcal{E}_{\rm bind}(n_0)$, $n_0$, $m_N^*(n_0)/m_N$, $K$, and $J$) are the same as in the original KVOR model. The other characteristics change the stronger, the closer to $f(n_0)$ the parameter $f_\om$ is chosen.  So, only one parameter changes slightly in the KVORcut04 model having $K_{\rm sym} \simeq -87$\,MeV and in the KVORcut03 model having $K_{\rm sym} \simeq -88$\,MeV, whereas $K_{\rm sym}\simeq -86$\,MeV in the KVOR model. In the KVORcut02 model two parameters change with $K_{\rm sym}\simeq -78$\,MeV and  $K'\simeq -869$\,MeV.

The KVOR model uses a sufficiently high value of  the effective nucleon mass at nuclear saturation $m^{*}(n_0)=0.805m_N$. In Ref.~\cite{Kolomeitsev:2004ff} it was shown that the smaller
$m^{*}(n_0)$ is in an RMF model, the larger the value of the maximum neutron star mass becomes. Thus, another way to increase $M_{\rm max}$ is to choose a smaller value of $m^*(n_0)$.
We formulate now the new model, labeled as MKVOR, which combines several mechanisms to stiffen the EoS discussed above.
In the MKVOR model the scaling functions are taken in Eq. (\ref{Efunct}) as follows:
\begin{align}
&\eta_\sigma(f) = \Big[1 - \frac{2}{3} C_\sigma^2 b f -
\frac{1}{2} C_\sigma^2 c_1 f^2 + \frac{1}{3} d f^3\Big]^{-1} \,,\quad c_1 = c -
\frac{8}{9} C_\sigma^2 b^2 \,,
\nonumber\\
&\eta_\omega(f) = \Big(\frac{1 + z \bar{f}_0}{1 + z f}\Big)^\alpha +
\frac{a_\om}{2} \left[1 + \tanh(b_\om (f - f_\om))\right]\,,
\label{KVORM_etar}\\
&\eta_\rho(f) = a_\rho^{(0)} + a_\rho^{(1)} f +
\frac{a_\rho^{(2)} f^2}{1 + a_\rho^{(3)} f^2}  +
\beta \exp\left(- \gamma \frac{(f - f_\rho)^2 ({1 + e_\rho
(f-\bar{f}_0)^2})}{1 + d_\rho
{(f-\bar{f}_0)}+  e_\rho (f-\bar{f}_0)^2} \right)\,.
\nonumber
\end{align}
As in the previously discussed models the hyperon and $\phi$ meson terms are omitted.
Parameters of the model are listed in Table~\ref{MKVORtabl}.

The form and the parameters of the scaling functions in MKVOR model are tuned to satisfy best the experimental constraints, that we demonstrate below, and to keep a connection to the KVOR model parametrization. Indeed, the first term in $\eta_\om$ is the same as in the KVOR model,
the function $\eta_\sigma$ and the first three terms in $\eta_\rho$ are basically the re-parametrization of the functions of the KVOR model. The new terms, the second one in $\eta_\om$ and the last one in $\eta_\rho$, are added to control the growth of the scalar field with an increase of the density. Other parameters of $\eta_\rho$ are also fine-tuned.

\begin{table}
\caption{Parameters of the MKVOR model and characteristics of the EoS at saturation.}
\begin{center}
\setlength{\tabcolsep}{4pt}
\begin{tabular}{cccccccccc}
\hline\hline
$C_\sigma^2$ & $C_\om^2$ & $C_\rho^2$ & $b \cdot 10^3$ & $c \cdot 10^3$ &  $d$ &$\alpha$ & $z$& $a_\om$ &$b_\om$  \\ \hline
 234.15 & 134.88 &  81.842 & 4.6750 & $-$2.9742 &  $-$0.5 & 0.4 & 0.65 & 0.11 &  7.1\\
\hline
$f_\om$& $\beta$& $\gamma$ & $f_\rho$ & $a_\rho^{(0)}$& $a_\rho^{(1)}$ & $a_\rho^{(2)}$ & $a_\rho^{(3)}$ & $d_\rho$ & $e_\rho$ \\
\hline
0.9 & 3.11 & 28.4 & 0.522 & 0.448 & $-$0.614 & 3 & 0.8 & $-$4 & 6 \\
\hline
\hline
 & $\mathcal{E}_0$ & $n_0$ & $K$ & $m_N^*(n_0)$ & ${J}$ & $L$ &$K'$ & $K_{\rm sym}$ &    \\
\cline{2-9}
 & [MeV] & [fm$^{-3}$] & [MeV] & $[m_N]$  & [MeV] & [MeV] &[MeV] & [MeV] &     \\
\cline{2-9}
 & $-16$ & 0.16 & 240 &  0.73 &  30 & 41 &  557 & $-$159 &     \\
\hline \hline
\end{tabular}
\end{center}\label{MKVORtabl}
\end{table}

\subsection{Scaling functions of KVORcut and MKVOR models}

\begin{figure}
\centering
\includegraphics[width = 11cm]{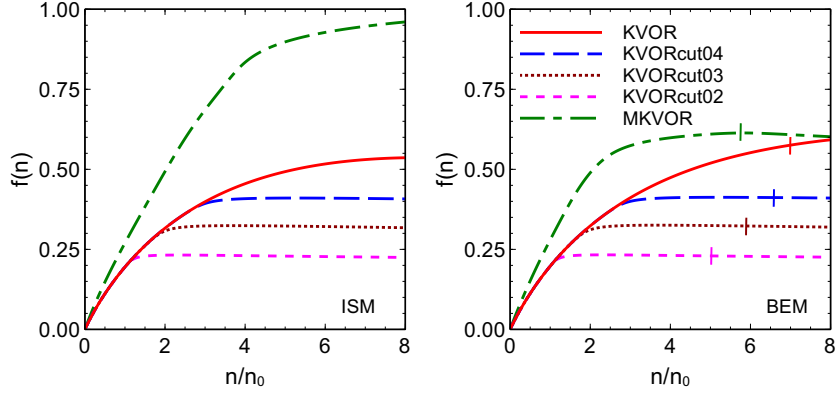}
\caption{The scalar field variable $f$ as a function of the nucleon density following as a solution of Eq.~(\ref{eq_fn}) in the ISM (left panel) and the BEM (right panel) for the KVOR, KVORcut and MKVOR models. Vertical bars on the right panel show the maximum densities reachable in stars with the maximum masses for the corresponding model.
} \label{fig:fn}
\end{figure}

\begin{figure}
\centering
\includegraphics[width = 13cm]{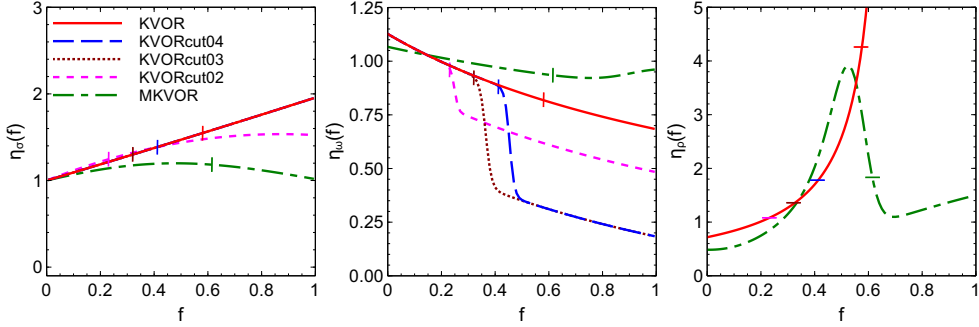}
\caption{Scaling functions $\eta_\sigma, \eta_\om, \eta_\rho$ as functions of the scalar field $f$ for KVOR, KVORcut, and MKVOR models. Vertical and horizontal bars indicate the maximum values of $f$ reachable in stars with the maximum masses for the corresponding model.}
	\label{fig:eta-f}
\end{figure}

\begin{figure}
\centering
\includegraphics[width = 13cm]{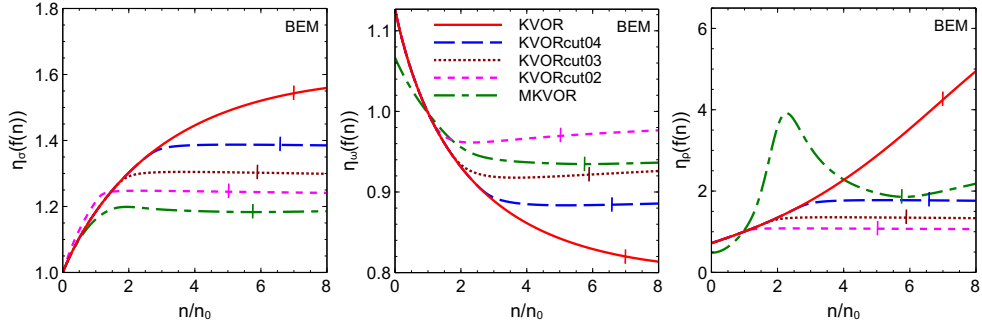}
\caption{Scaling functions $\eta_\sigma, \eta_\om, \eta_\rho$ as
functions of the density following from Eq.~(\ref{eq_fn}) for the BEM. Vertical bars indicate central densities in the stars with the maximum masses for the corresponding model.}
\label{fig:eta-n}
\end{figure}

The scalar field variables $f(n)$ for the KVOR, three KVORcut and MKVOR models are shown in Fig. \ref{fig:fn} for the ISM (left panel) and for BEM (right panel). In all models, except
MKVOR in the ISM, the function $f(n)$ saturates at high densities at the values smaller than 1.
The dependence of the scaling functions $\eta_{m}$ on $f$ is illustrated in Fig.~\ref{fig:eta-f}. We see that $\eta_\om (f)$ changes abruptly at a certain value of $f$ for the KVORcut models that allows to stabilize the function $f(n)$ at a certain value. For the MKVOR model the same mechanism of the $f$ stabilization is implemented not in the $\eta_\om (f)$ function but in the $\eta_\rho(f)$, which drops fast for $f> 0.52$. The rise of $\eta_\rho(f)$ for $f<0.5$ serves to tame the growth of the nuclear symmetry energy with an increasing density and to provide, thereby, a sufficiently high DU threshold. Note that in the KVOR model the $\eta_\rho (f)$ diverges at $f\simeq 0.7$ but such a large value of $f$ is not realized in the neutron star within the KVOR model. How one can formally avoid this divergence without a change of $\eta_\rho (f)$ in the physical region of parameters was demonstrated in Ref.~\cite{Khvorostukhin:2006ih} and explained above.
In Fig.~\ref{fig:eta-n} we depict $\eta$ scaling functions as in Fig.~\ref{fig:eta-f} but now computed for $f=f(n)$ for the BEM, as follows from the solution of Eq.~(\ref{eq_fn}).

\begin{figure}
\centering
\includegraphics[width = 11cm]{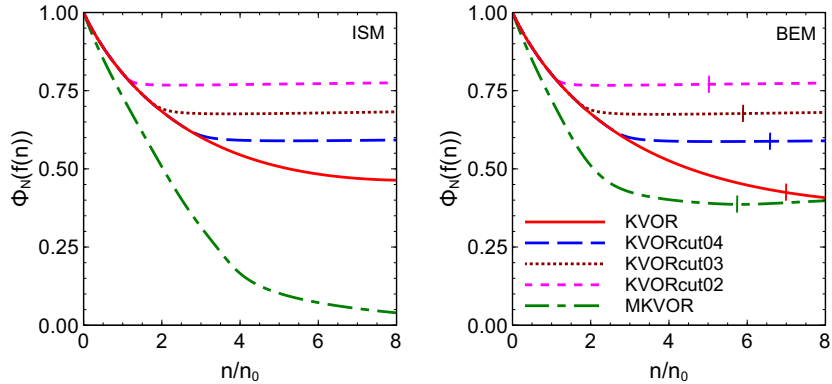}
\caption{The common scaling function for nucleon and meson masses as a function of the density
calculated according Eq.~(\ref{PhiN}) with $f(n)$ from Eq.~(\ref{eq_fn})
for ISM (left panel) and  BEM (right panel) for various models. }\label{fig:Phi-n}
\end{figure}

The scaling functions of the effective nucleon mass $\Phi_N$ as functions of the nucleon density for various models are shown in Fig.~\ref{fig:Phi-n} for the ISM (left panel) and the BEM
(right panel). According to our assumption the same function controls also the variation of
$\sigma$, $\omega$ and $\rho$ meson masses. In all models the effective hadron masses decrease first with the density and then saturate at some constant values, except the KVOR model and MKVOR model in the ISM. The limiting values of the effective hadron masses at $n\gg n_0$ (here presented up to $8\,n_0$) in ISM are 0.78, 0.68, 0.59 for KVORcut02, KVORcut03 and
KVORcut04 models, respectively, and in BEM they are 0.78, 0.68,
0.59, 0.39 for KVORcut02, KVORcut03, KVORcut04 and MKVOR models,
respectively.
Note that the behavior $m^{*}_N (n)$ demonstrated in Fig.~\ref{fig:Phi-n} is qualitatively in a
line with that expected within the renormalization group approach in Ref.~\cite{Paeng:2013xya}, where the effective nucleon mass undergoes an almost linear drop for densities up to a value above $n_0$ and then stays approximately constant.

\begin{figure}
\centering
\includegraphics[width = 13cm]{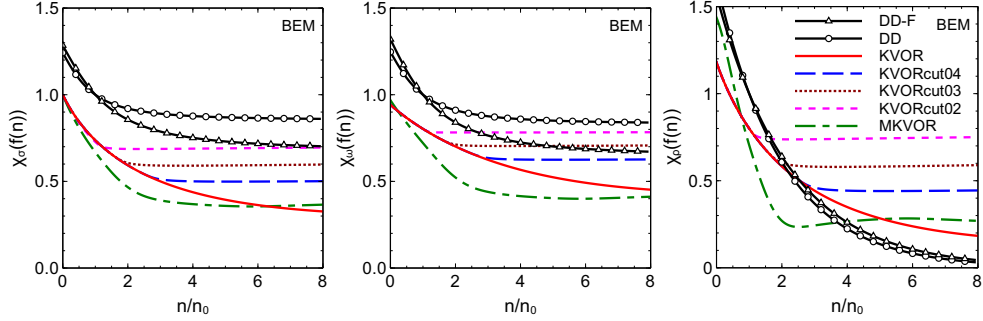}
\caption{The scaling functions for the coupling constants $\chi_m$, $m=\sigma$, $\om$, $\rho$, as functions of the total baryon density in the BEM for the KVOR, KVORcut and MKVOR models. Solid lines with circles and triangles depict the density dependence of the coupling constants in the DD~\cite{Typel2005} and DD-F~\cite{Klahn:2006ir} models, respectively.
    }\label{scalchi}
\end{figure}

The scaling functions of the coupling constants $\chi_m$  are shown in Fig.~\ref{scalchi} as functions of the density in BEM for our models in comparison with those in the DD~\cite{Typel2005} and DD-F~\cite{Klahn:2006ir} models, where hadron coupling constants are assumed to be  density dependent, but meson masses stay constant, $\Phi_m =1$. We see  that the density dependencies of our scaling functions for the $\sigma N$, $\om N$ and $\rho N$ coupling constants prove to be similar to those exploited in the DD and DD-F models.
For $n\lsim (2\mbox{--}3)\,n_0$ in the KVORcut and MKVOR models and for all $n$ in the KVOR model the $\rho NN$ coupling constant decreases more rapidly with the density, than the $\omega NN$ coupling constant, that is qualitatively similar to the behavior discussed in Ref.~\cite{Paeng:2013xya}.

It is instructive to compare the scalar, $S_N$, and vector, $V_N$, potentials acting on a nucleon in ISM
\begin{eqnarray}
S_N(n)=-m_N\,f(n) \,,\quad  V_N=\frac{C_\om^2\,n}{2\,m_N\,\eta_\om(f(n))}
\label{UN-pot}
\end{eqnarray}
 with the results of the Dirac-Brueckner-Hartree-Fock (DBHF) calculations~\cite{BrockMarch,KatSaito}. We see that for the MKVOR model the potentials are comparable with those follow from the new microscopic calculations ~\cite{KatSaito}, but  the former rise with a density faster  at small densities than the potentials calculated in~\cite{BrockMarch}.

\begin{figure}
\centering
\includegraphics[width = 6cm]{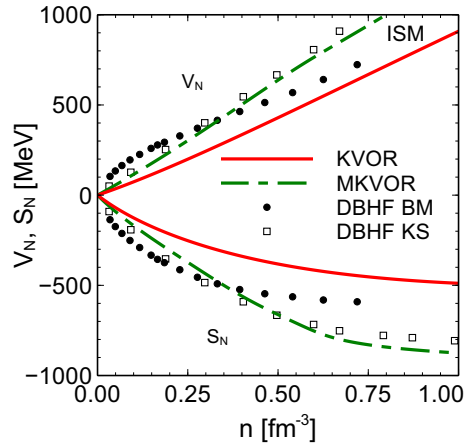}
\caption{
The scalar and vector potentials, Eq.~(\ref{UN-pot}), acting on a nucleon in ISM as functions of nucleon density for KVOR and MKVOR potentials in comparison with the DBHF calculations~\cite{BrockMarch,KatSaito} for the Bonn A potential.
}
\label{dbhf-u}
\end{figure}

\section{Energy per nucleon, Landau parameters and speed of sound}
\label{Thermochar}

\begin{figure}
\centering
\includegraphics[width = 11cm]{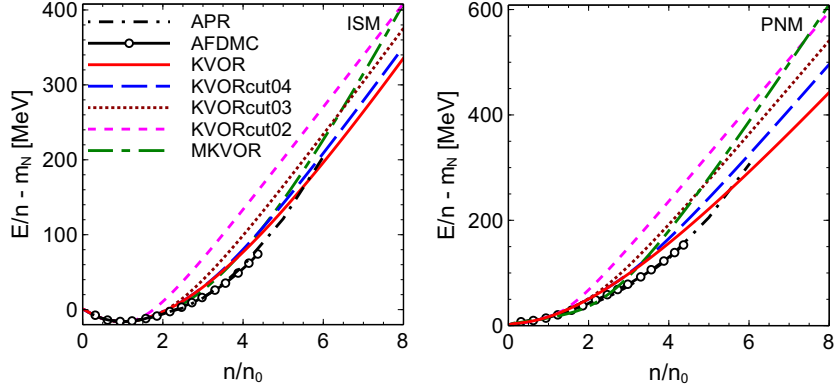}
\caption{Energy per particle as a function of the nucleon density
for ISM (left panel) and for PNM (right panel). Short
dash-dotted line shows the APR EoS~\cite{APR} and the solid line
with circles stands for the AFDMC simulations~\cite{Gandolfi:2009nq}.
}
\label{EnergySymPNM}
\end{figure}

The energy per particle as a function of the nucleon density is shown in Fig.~\ref{EnergySymPNM}    for the ISM (left panel) and for the purely neutron matter (PNM) (right panel) for the KVOR,
KVORcut and MKVOR models in comparison with the microscopic APR
EoS~\cite{APR} and the auxiliary field diffusive  Monte Carlo (AFDMC) calculations~\cite{Gandolfi:2009nq}. All considered EoSs demonstrate a similar behavior
for $n\lsim 1.5 n_0$ and begin to deviate substantially from each other demonstrating different stiffness with a density increase.

The energy of a nucleon in the nuclear matter with proton and neutron particle distributions
$f_p(\vec{p}\,)$ and $f_n(\vec{p}\,)$ follows from the Lagrangian in the mean-field approximation (\ref{Lag-bar}) as
\begin{align}
\varepsilon_i(\vec{p}\,)=\sqrt{m_N^2\Phi_N^2(\bar{f})+(\vec{p}-\vec{V}_i(\bar{f}))^2}
+V_{0,i}(\bar{f})\,,\quad i=n,p\,,
\label{NuclFL-eN}
\end{align}
where vector potentials and currents are defined by the relations
\begin{align}
&\frac{V_{0,p}(f)+V_{0,n}(f)}{2}=V^{(\om)}_{0}(f)=\frac{C_\om^2(n_n+n_p)}{m_N^2\eta_\om(\bar{f})}
\,,
\nonumber\\
& \frac{V_{0,p}(f) - V_{0,n}(f)}{2}=V^{(\rho)}_0(f)=\frac{C_\rho^2 (n_p-n_n)}{4m_N^2\eta_\rho(\bar{f})}\,,
\nonumber\\
&\frac{\vec{V}_{p}(f) + \vec{V}_n(f)}{2} =\frac{C_\om^2(\vec{j}_n+\vec{j}_p)}{m_N^2\eta_\om(\bar{f})}\,,
\quad
\frac{\vec{V}_{p}(f) - \vec{V}_n(f)}{2}=\frac{C_\rho^2 (\vec{j}_p-\vec{j}_n)}{4m_N^2\eta_\rho(f)}\,,
\nonumber\\
& n_i=\intop \frac{2\,{\rm d}^3 p}{(2\pi)^3}\, f_i(\vec{p}\,)\,,\qquad
\vec{j}_i=\intop \frac{2\,{\rm d}^3p} {(2\pi)^3}\,
\frac{\vec{p}-\vec{V}_i(\bar{f})}{\sqrt{(\vec{p}-\vec{V}_i(\bar{f}))^2+m_N^2\Phi_N^2(\bar{f})}}\,
f_i(\vec{p})\,.
\label{NuclFL-pot}
\end{align}
The scalar field parameter $\bar{f}$ follows from Eq.~(\ref{eq_fn})
\begin{align}
&\frac{m_N^4 \bar{f}^2}{2C_\sigma^2}\eta_\sigma(\bar{f}) \Big[\frac{2}{\bar{f}}+\frac{\eta'_\sigma(\bar{f})}{\eta_\sigma(\bar{f})}\Big]
+ V^{(\om)}_{0}(\bar{f})(n_n+n_p)\frac{\eta'_\om(\bar{f})}{\eta_\om(\bar{f})}
+ V^{(\rho)}_0(\bar{f})(n_n-n_p)\frac{\eta'_\rho(\bar{f})}{\eta_\rho(\bar{f})}
\nonumber\\
&\qquad\qquad
=-m_N^2\Phi_N(\bar{f})\Phi'_N(\bar{f})
\intop \frac{2\,{\rm d}^3 p}{(2\pi)^3}\, \frac{f_p(\vec{p}\,)+f_n(\vec{p}\,)}{\sqrt{(\vec{p}-\vec{V}_i(\bar{f}))^2+m_N^2\Phi_N^2(\bar{f})}}\,.
\label{LP-feq}
\end{align}

We consider the nucleon distributions with a slightly distorted,  non-spherical Fermi-surface, so that the nucleon current is non-zero, $\vec{j}_i\neq 0$. For example we can take
$f_i(\vec{p})=\theta(p_{\rmF,i}[1+\epsilon/3+\vec{a}\vec{p}/p_{\rmF,i}+\dots]-p)$, where  $\epsilon\ll 1$, $|\vec{a}|\ll 1$ are small parameters.
The variation of the nucleon energy with respect to nucleon distribution defines the Landau parameters. For momenta close to the Fermi surface we can write, cf.~\cite{Matsui},
\begin{align}
F_{ij}(\vec{n},\vec{n}\,')=\frac{\delta \varepsilon_{i}(p_{\rmF,i}\vec{n})}
{n_j\delta f_{j}(p_{\rmF,j}\vec{n}\,')}\,.
\label{LP-def}
\end{align}
We take here into account that the variation of the density under the variation of the particle distribution is $\delta n_j=n_j\delta f_j\approx n_j\epsilon$.

The Landau parameters $F_{ij}$ can be expressed as  matrix elements of the effective nucleon-nucleon interaction matrix in the particle-hole channel,
\begin{align}
F_{ij}(\vec{n}\,',\vec{n}\,)=(\chi^\dag_j)_\alpha (\chi_i)_\gamma \widehat{\mathcal{F}}_{\alpha\gamma,\beta\delta}(\vec{n}\,',\vec{n}\,)
(\chi^\dag_i)_\beta (\chi_j)_\delta\,,
\label{LP-matel}
\end{align}
where $\chi_i$ is the isospin spinor of the nucleon of type $i$, $(\chi_p)_\alpha=\delta_{1\alpha}$ and
$(\chi_n)_\alpha=\delta_{2\alpha}$\,, and the interaction matrix can be written in terms of the Pauli matrices acting in the nucleon isospin state,
\begin{align}
\widehat{\mathcal{F}}_{\alpha \gamma, \beta \delta}(\vec{n}\,' , \vec{n}\,)=
F(\theta)\,\delta_{\alpha \delta} \delta_{\beta\gamma} +
F'(\theta)\, \vec{\tau}_{\alpha \delta} \vec{\tau}_{\beta\gamma}\,.
\label{Fteta}
\end{align}
Here $F$ and $F'$ are functions of the angle  between the directions of the Fermi momenta of incoming and outgoing nucleons, $\cos\theta=(\vec{n}\,' \vec{n}\,)$.
From Eqs.~(\ref{LP-matel},\ref{Fteta}) we find  relations, cf.~\cite{Migdal-TKFS}
\begin{align}
F_{nn}=F_{pp} = \widehat{\mathcal{F}}_{22,22} =F+F'
\,,\quad
F_{np} = \widehat{\mathcal{F}}_{12,21} =F-F'\,.
\label{Fnn-Fnp-relat}
\end{align}
The Landau parameters are usually expanded in the Legendre polynomials
\begin{eqnarray}
F_l = (2 l + 1) \int \frac{d \Omega}{4 \pi} P_l (\cos \theta) F(\theta), \,\,
F'_l = (2 l + 1)  \int \frac{d \Omega}{4 \pi} P_l (\cos \theta)
 F'(\theta)\,.\label{fl}
\end{eqnarray}
The Landau parameters $F_0$ and $F_1$ are important
characteristics of the nuclear Fermi liquid. For the Walecka model
they were first calculated in Ref.~\cite{Matsui}. Following~\cite{Matsui}, we define the dimensionless Landau parameters $f_{0,1}$ as follows:
\begin{align}
f_i = N_\rmF F_i, i = 0,1, \quad
N_\rmF = \gamma p_\rmF E_\rmF/(2 \pi^2), \quad
p_\rmF = (6 \pi^2 n/\gamma)^{1/3}.
\label{normal}
\end{align}
$N_\rmF$ is the density of states at the Fermi surface,
$E_\rmF=\sqrt{p_\rmF^2+m_N^{*\,2}}$ is the relativistic Fermi
energy of the nucleon. For ISM $\gamma = 4$, for the PNM $\gamma
=2$. We will consider here these two cases, since for the nuclear matter of arbitrary isotopic composition expressions are more cumbersome.

The incompressibility $K$, the square of the  sound
velocity $v_s^2$ and the chemical potential in the ISM are expressed through $f_0(n)$ as
\begin{align}
& K = \frac{3 p_\rmF^2}{E_\rmF}(1 + f_0),
\,\,\,\,
v_s^2  = \frac{p_\rmF^2}{3 E_\rmF \mu} (1 + f_0), \,\,\,\,
\mu = \frac{C_\om^2 n}{m_N^2 \eta_\om(f)} + \sqrt{p_\rmF^2 + m_N^{*2}}.
\label{lp_K}
\end{align}
Note that in the relativistic theory the particle energy at the Fermi surface plays the same role as the effective mass in the non-relativistic Fermi-liquid theory. It can be expressed in terms of the Landau parameter $f_1$:
\begin{align}
E_{\rm F} = \sqrt{p_{\rm F}^2 + m_N^{*2}}=\mu (1 + \frac{1}{3} f_1).
\label{lp_ef}
\end{align}
The symmetry energy is connected with $f'_0$ in the ISM by the following relation:
\begin{align}
\widetilde{\cal{E}}_{\rm sym} = \frac{p_\rmF^2}{6 E_\rmF}(1 + f'_0) .
\label{lp_j}
\end{align}
For our generalized RMF models with scaled hadron masses and coupling constants the Fermi liquid
parameters in ISM and in PNM are presented in Appendix~\ref{app:Landpar}.

\begin{figure}
\centering
\includegraphics[width = 14cm]{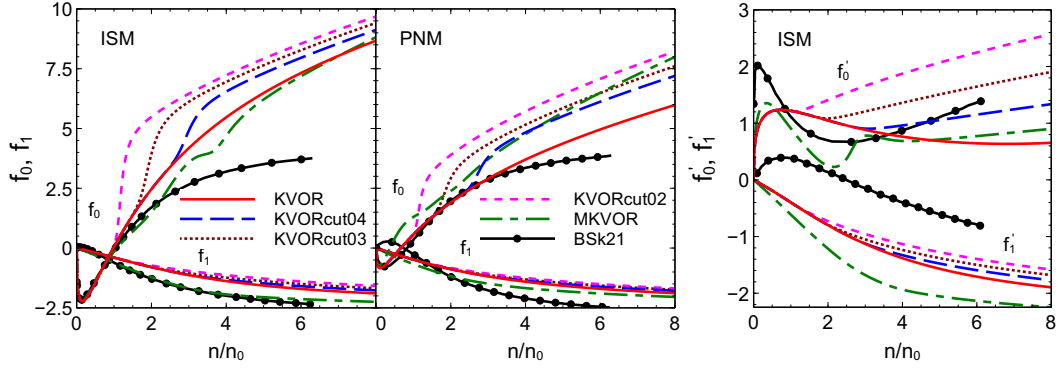}
\caption{The density dependence of the dimensionless Landau
parameters $f_{0}$ and $f_1$ in the ISM and the PNM, and of the parameters $f'_0$ and $f'_1$ in the ISM. Solid curves with filled dots show calculations within the Skyrme BSk21 model~\cite{Goriely:2010bm}. Notations of other curves are the same as in previous Figs.
 }
 \label{fig:LP}
\end{figure}

In Fig.~\ref{fig:LP} we demonstrate the density dependence of the
Landau parameters $f_0$ and $f_1$ computed within the KVOR, KVORcut and MKVOR models. The results are shown for the ISM (left panel) and the PNM (right panel) in comparison with calculations performed in the framework of the Skyrme model BSk21~\cite{Goriely:2010bm}. We see that the Landau parameters in all models demonstrate the qualitatively similar behavior. In the ISM for $5\times 10^{-3}~n_0 \lsim n\lsim 0.6~n_0$ we get the scalar Landau parameter $f_0<-1$. In this density interval there appears the Pomeranchuk instability with respect to the growth of the long-wave density fluctuations. In the PNM $f_0>-1$ and such an instability does not arise.

\begin{figure}
\centering
\includegraphics[width = 10cm]{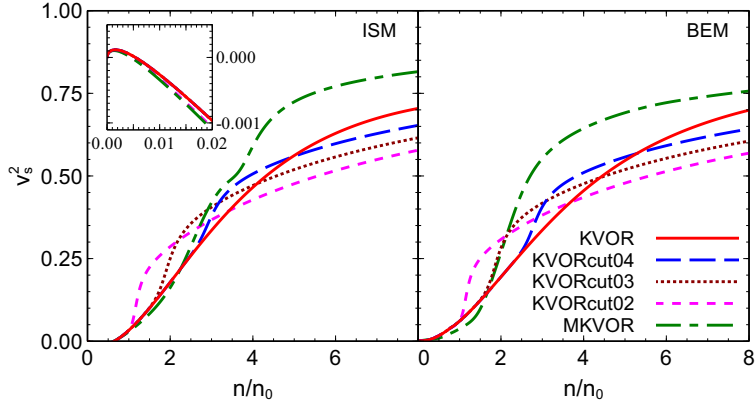}
\caption{The squared speed of sound as a function of the
nucleon density for various RMF models in the ISM (left panel) and in the BEM (right panel).
}
\label{fig:vs}
\end{figure}

In Fig.~\ref{fig:vs} we show the behavior of the squared speed of sound as a function of the nucleon density in the ISM (left panel) and in the BEM (right panel). The sound velocity
demonstrates a monotonous rise with a density increase but  never reaches the velocity of the light ($c=1$). We see also that $v_s^2$ becomes negative in the density interval, where the scalar Landau parameter $f_0<-1$, i.e. in the region of the Pomeranchuk instability, see \cite{KV2015}.

In the BEM the squared speed of sound bends slightly (not seen visually on the plot) at $n \sim 0.8 n_0 $,  which is the point of the muon appearance. Such a behavior is typical for third-order phase transitions.

\section{Comparison of the EoS characteristics with experimental constraints}
\label{ExpConstraints}

\subsection{Optical potential}

The baryon optical potential in the ISM is determined as~\cite{Feldmeier,Delfino}
\begin{align}
U_{{\rm opt}}^{b}(\varepsilon) = \varepsilon -\sqrt{(\varepsilon - V_b)^2 - S_b\, (2\, m_b + S_b)},
\label{U-opt}
\end{align}
where $\varepsilon$ is the baryon energy, the scalar and vector potentials acting on the given baryon in ISM are defined as
\begin{align}
S_b = m_b \Phi_b(x_{\sigma b} \frac{m_N}{m_b} f) - m_b\,,
\quad V_b = x_{\om b} \frac{C_\om^2\, n}{m_N^2 \eta_\om(f)}\,.
\label{SV-pot}
\end{align}
The expression for the optical potential of antibaryons is obtained by replacing $V_b \rightarrow -V_b$.

\begin{figure}
\centering
\includegraphics[width = 11cm]{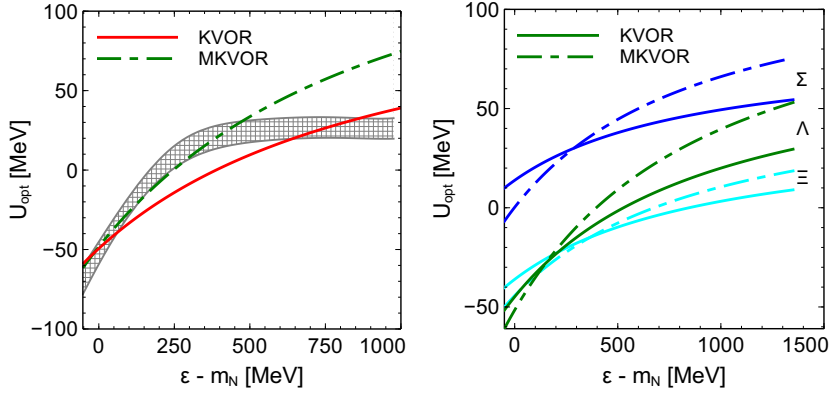}
\caption{Left panel: Energy dependence of the nucleon optical potential in the ISM at $n=n_0$ calculated for the KVOR and MKVOR models. Hatched area shows the extrapolation from finite nuclei to the nuclear matter~\cite{Feldmeier}. Right panel: Optical potentials of $\Sigma$, $\Lambda$ and $\Xi$ for the same models.
} \label{fig:Uopt-B}
\end{figure}

The dependence of the nucleon optical potential on the nucleon energy in the ISM at $n = n_0$ is shown in Fig.~\ref{fig:Uopt-B} (left panel). The hatched band is the optical potential extracted from the atomic nucleus data~\cite{Hama} and recalculated to the case of the infinite nuclear ISM in Ref.~\cite{Feldmeier}. We see that the MKVOR model describes the nucleon optical potential rather appropriately for energies $\varepsilon-m_N \lsim 400$ MeV. To match the data for higher particle energies a momentum dependence of the $NN$ interaction would be required, which is not present in the mean-field approach. The KVOR model describes the nucleon optical potential for low and high energies better than the MKVOR model but does not describe it for
intermediate energies. Calculations for the KVORcut models do not differ from that for KVOR since for $n=n_0$ differences in the relevant parameters of these models are minor.
The isovector part of the optical potential $U_{\rm opt}^n(\varepsilon)- U_{\rm opt}^p(\varepsilon)$ is less constrained by the data and we do not discuss it, therefore.

\begin{figure}
\centering
\includegraphics[width = 11cm]{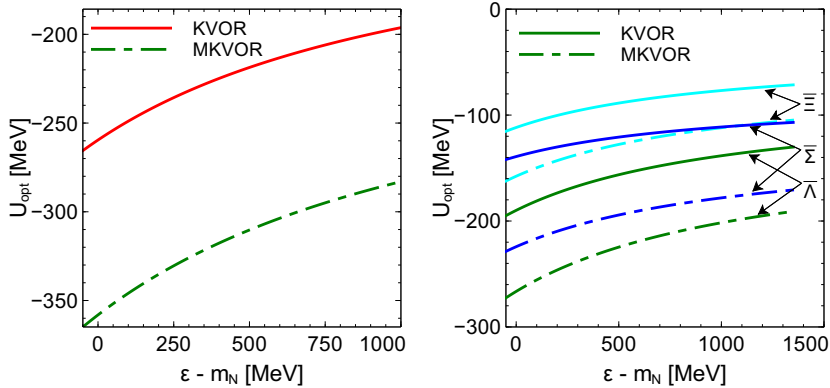}
\caption{Energy dependence of the antinucleon optical potential (left panel) and of the antihyperon one (right panel) for the  ISM at $n=n_0$. }
\label{fig:Uopt-Bbar}
\end{figure}

The energy dependence of the hyperon optical potentials is not yet constrained by data.
Predictions of our models for hyperons are shown in Fig.~\ref{fig:Uopt-B}\,(right panel).  Predictions for the optical potentials $U_{\rm opt}^{\bar{b}}(\varepsilon)$ of antinucleons and antihyperons in the ISM at $n=n_0$ are depicted in Fig.~\ref{fig:Uopt-Bbar}. Predictions for antiprotons are very important in view of the future experiments at FAIR. For the
KVOR model the antinucleon optical potential was calculated in Ref.~\cite{Khvorostukhin:2008xn}. A phenomenological value of the antiproton optical potential is limited within the range
$-(100\mbox{---}350)$\,MeV~\cite{WangKerman}. The available experimental
data from $\bar{p}$ atoms~\cite{FGM} and $\bar{p}$ scattering off nuclei~\cite{Walker} suggest that the depth of the real part of the $\bar{p}$--nucleus potential in the interior of a nucleus lies in the range $-(100\mbox{---}300)$\,MeV.

\subsection{Low density behavior of EoS}

The behavior of the nucleon EoS for $n<n_0$ was extensively studied within various phenomenological and microscopic models including the chiral effective field theory.
The EoS derived from our RMF model can be, of course, tuned to one of realistic parameterizations
of the EoS for $n<n_0$ and vanishing temperature, but this is not our goal in the present work. We just continue to exploit simple analytical parameterizations of the scaling factors $\eta_{\sigma}$, $\eta_{\om}$ and $\eta_{\rho}$ presented above.

\begin{figure}
\centering
\includegraphics[width = 11cm]{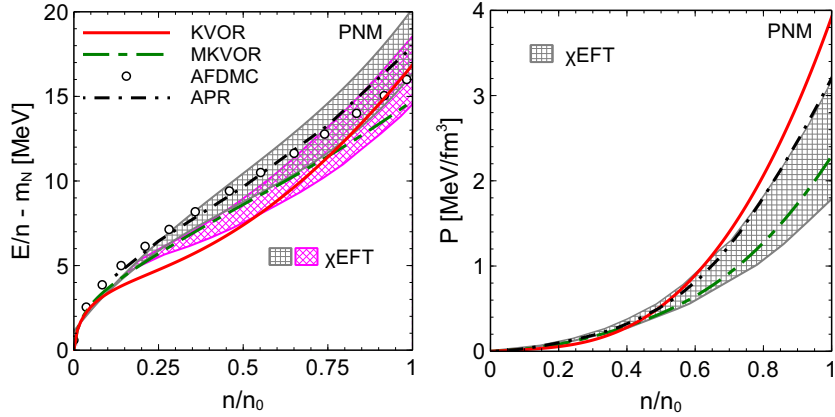}
\caption{Energy per particle (left) and pressure (right) for PNM
as functions of the nucleon density for $n<n_0$. Overlapping bands
are evaluated  from chiral effective field theory for different
$NN$ potentials and include uncertainty estimates due to the
many-body effects~\cite{Hebeler:2014ema}. AFDMC
calculations~\cite{Gandolfi:2009nq} are shown by empty circles,
and short dash-dotted line stands for the APR EoS. Solid line presents
calculation with the KVOR EoS, long dash-dotted line is for MKVOR
EoS.
 }
 \label{fig:EnergyLown}
\end{figure}

The energy per particle and the pressure for the PNM are shown in Fig.~\ref{fig:EnergyLown} as functions of the nucleon density for $n<n_0$. The curves for all three KVORcut models coincide with the KVOR model in this density range. The MKVOR
curves lie within the uncertainty region estimated within the chiral effective field theory~\cite{Hebeler:2014ema}. The pressure of the KVOR model goes beyond the estimated uncertainty region. The difference between the energies per particle for the
AFDMC EoS~\cite{Gandolfi:2009nq}
and our MKVOR parameterization does not exceed 2 MeV for $n\lsim
n_0$. In comparison with APR EoS  A18$+\delta v+$UIX$^{*}$ the difference is a bit larger.

\begin{figure}
\centering
\includegraphics[width = 11cm]{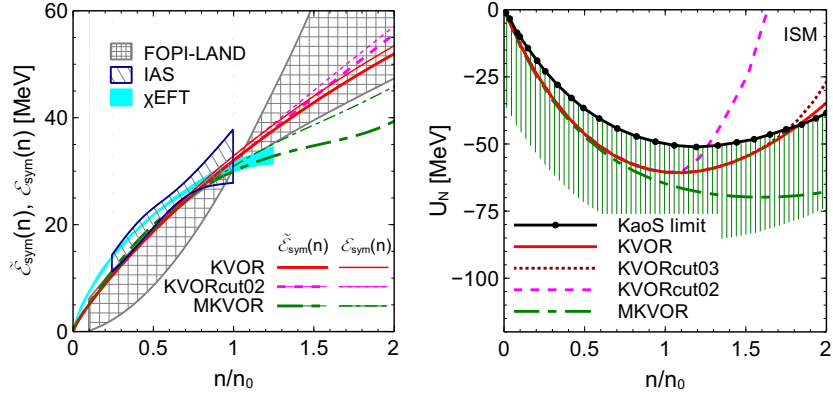}
\caption{Left panel: The symmetry energy coefficients $\widetilde{\mathcal{E}}_{\rm sym}(n)$, cf. Eq.~(\ref{Jtilde}), and $\mathcal{E}_{\rm sym}(n)$ given by Eq.~(\ref{Jn}) as functions of the nucleon density calculated for various models. The bands are constraints from chiral effective field theory  ($\chi$EFT)~\cite{Hebeler:2014ema}, from study of
IAS~\cite{DanielewiczLee}, and from the FOPI-LAND experimental data for Au+Au collisions at 400\,MeV/nucleon~\cite{Cozma:2013sja}.
Right panel: The nucleon potential $U_N$ as a function of the nucleon density in the ISM for various models.
The curve labeled as KaoS shows the upper boundary for the potential deduced~\cite{Sagert:2011kf} from the kaon production measured by KaoS collaboration.
 }
 \label{fig:JLown}
\end{figure}

Constraints from Ref.~\cite{Cozma:2013sja} on the density dependence of the
symmetry energy $\widetilde{\mathcal{E}}_{\rm sym}(n)$, cf. Eq.~(\ref{Jtilde}), were deduced from the comparison of theoretical predictions for the difference in proton and neutron flows with the experimental data of FOPI-LAND collaboration for Au+Au collisions at 400\,MeV/nucleon. The constraints are shown in Fig.~\ref{fig:JLown} together with the constraints from the chiral effective field theory~\cite{Hebeler:2014ema} and from the study of the isobaric analog states (IAS)~\cite{DanielewiczLee}.
In Fig.~\ref{fig:JLown} we plot the symmetry energy derived for our KVOR, KVORcut02 and MKVOR models. The curves for KVORcut04 and KVORcut03 models coincide with the KVOR curve.
Note that the KVOR and KVORcut models yield $J=32$\,MeV, whereas the MKVOR model corresponds to $J=30$\,MeV. Because of this the MKVOR curve lies
below the KVOR one at $n_0$. The value $\mathcal{E}_{\rm sym}(n)$ is given by Eq.~(\ref{Jn}). For $n<n_0$ the difference between $\widetilde{\mathcal{E}}_{\rm sym}(n)$ and $\mathcal{E}(n)$ is minor. It increases however with a density increase for $n>n_0$ and $\widetilde{\mathcal{E}}_{\rm sym}(n)<\mathcal{E}_{\rm sym}(n)$. For $n> n_0$ curve $\mathcal{E}_{\rm sym}(n)$ for the MKVOR model (thin dash-dotted line) deviates slightly from the lower boundary of region estimated from the FOPI-LAND data, whereas the deviation is larger for the quantity $\widetilde{\mathcal{E}}_{\rm sym}(n)$ (thick dash-dotted line). The lines for the KVOR and KVORcut models lie inside the region.

The nucleon potential, $U_N(n)= S_N+V_N$ with the scalar $S$ and vector potential $V$ defined in Eq.~(\ref{SV-pot}), can also be constrained in the ISM at low nucleon densities from measurements of the kaon production in heavy-ion collisions performed by the Kaon Spectrometer (KaoS)~\cite{Sagert:2011kf}. It is shown on the right panel in Fig.~~\ref{fig:JLown} together with the curves produced by our models, see also~\cite{Lopes:2013cpa}. We find that KVORcut02 model is too stiff to fulfill the constraint. The KVOR, KVORcut04 and
KVORcut03 models fulfill the constraint except near the border $2n_0$.
The model MKVOR fulfills the constraint fully.

\subsection{Constraint from particle flow in heavy-ion collisions}

\begin{figure}\centering
\includegraphics[width = 11cm]{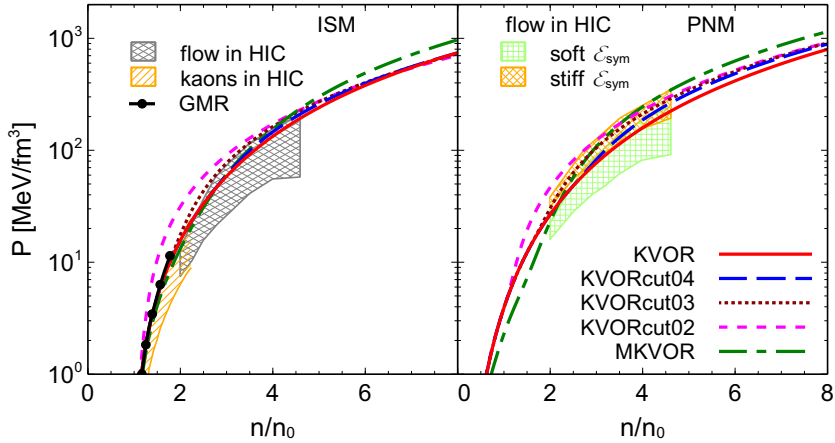}
\caption{  Pressure as a function of the nucleon density for ISM (left panel) and for the PNM (right panel). Double-hatched area is the constraint from the particle flow in heavy-ion collisions~\cite{Danielewicz:2002pu}, hatched area is the kaon flow constraint extracted in Ref~\cite{Lynch} from the analysis of Ref.~\cite{Fuchs1}.  Line with bold dots shows the extrapolation of the pressure consistent with the  GMR, cf. Ref.~\cite{Lynch}. On the right panel two double-hatched areas show the pressure  consistent with the flow data after inclusion of the isospin asymmetry terms with stiff and soft density dependencies.}
\label{PressureDaniel}
\end{figure}

The analysis~\cite{Danielewicz:2002pu} of experimental data on transverse and elliptical flows of particles in heavy-ion collisions reveals a correlation among the flow magnitude and the stiffness of the EoS in the ISM. The pressure as a function of the nucleon density in the ISM is shown in Fig.~\ref{PressureDaniel}\,(left panel). Double-hatched band ranging from $2n_0$ to about $4.5n_0$ corresponds to the so-called particle-flow constraint of Ref.~\cite{Danielewicz:2002pu}. It rules out very stiff EOSs and puts in a challenge the requirement that EoS should be stiff enough to fulfill the constraint on the maximum neutron star mass extracted from the measurement of
pulsar masses in Ref.~\cite{Demorest:2010bx,Antoniadis:2013pzd}. In Fig.~\ref{PressureDaniel} we show also a restriction on the stiffness of the EoS from $n\simeq 1.2 n_0$ to $n\simeq 2.2 n_0$
extracted in ~\cite{Lynch} from the analysis of the kaon flow in Ref.~\cite{Fuchs1} and the extrapolation of the pressures consistent with the GMR data analysis~\cite{Lynch}. The KVOR model satisfies the requirements from heavy-ion collisions. However, the maximum neutron star mass with the KVOR EoS is equal to
$2.01 M_\odot$ and fits only marginally the experimental value of the mass $2.01\pm0.04 M_{\odot}$ of the pulsar PSR J0348+0432~\cite{Antoniadis:2013pzd}. Since hyperons
are not included in the KVOR model and their inclusion may only result in a decrease of $M_{\rm max}$, the KVOR model needs a revision. The EoS for the KVORcut04 model proves to be
only slightly stiffer than that of the KVOR one: the pressure curve for this model passes very close to the curve for KVOR model. The curve for the KVORcut03 model goes over the upper boundaries of the hatched regions. The curve for the MKVOR model passes through the hatched regions for $n<4 n_0$ but escapes it at higher densities. Contrarily, the curve for the KVORcut02 model lies above the region for $n<3.5 n_0$ but enters the double-hatched
region for higher $n$. Thus, we see that the KVORcut03 and MKVOR
models are the most promising models for the simultaneous fulfillment of the
particle-flow and maximum neutron star mass constraints.

For the PNM the upper and lower double-hatched regions in Fig.~\ref{PressureDaniel}\,(right panel) correspond to the pressure in the ISM consistent with the flow data after inclusion of the pressure from the symmetry energy term, $\mathcal{E}_{\rm sym}(n)$, with a strong or weak density dependence. All presented curves except the KVORcut02 one fit the double-hatched region with the strong density dependence of $\mathcal{E}_{\rm sym}(n)$. The curve for KVORcut02 model lies slightly above the upper boundary of the double-hatched region in the region $n<3n_0$ that might indicate  too high stiffness of this EoS at these densities.

\subsection{Direct Urca constraint}

The DU processes on neutrons, $n \to p + e +\bar{\nu}_e$, $p+e\to n+\nu_e$, can occur if the Fermi momenta of particles satisfy the  inequality $p_{{\rm F},n} \leq p_{{\rm F},p}+p_{{\rm F},e}$. For the hyperon free matter this inequality can be rewritten with the help of the electro-neutrality condition~(\ref{electroneut}) as
\begin{align}
\frac{n_p}{n} > x_{\rm DU} =\frac{1}{1+(1+x_e^{1/3})^3}\,, \quad x_e = \frac{n_e}{n_e + n_{\mu}}\,,
\end{align}
where $x_{\rm DU}$ is the DU-threshold proton fraction. The values of the DU threshold density, $n_{\rm DU}^{(n)}$, and the corresponding mass of the star, $M^{(n)}_{\rm DU}$, strongly depend on the density dependence of the symmetry energy.

\begin{figure}\centering
\includegraphics[width = 11cm]{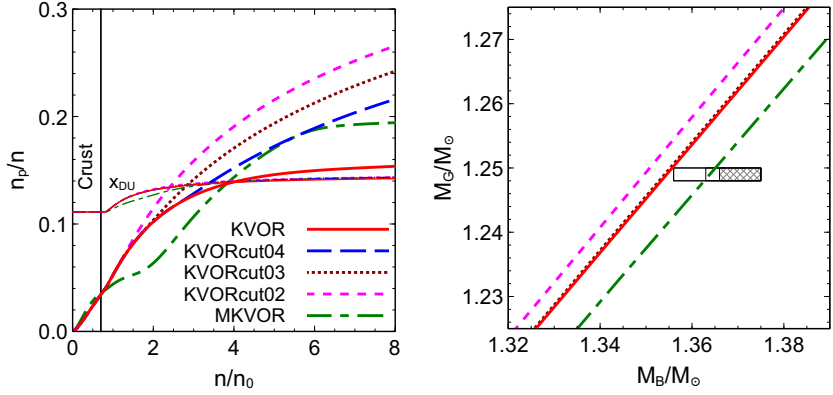}
\caption{ Left panel: Proton fractions in the BEM for various models (thick lines) and  the thresholds of the DU reaction $x_{\rm DU}$ (corresponding thin lines) as functions of the nucleon density. Vertical line indicates the upper border of the crust.
Right panel: Gravitational mass of a neutron star, $M_G$, versus the baryon mass, $M_B$, for various models.
The double-hatched rectangle is the constraint for the pulsar J0737-3039(B)~\cite{Podsiadlowski}. The two empty rectangles show the variation of the constraint, when the assumed loss of the baryon mass during the progenitor-star collapse amounts to $0.3\%\,M_{\odot}$ and $1\%\,M_{\odot}$.
Line for the KVORcut04 model coincides with that for the KVOR model. }
\label{Pods}
\end{figure}

In Fig.~\ref{Pods}\,(left panel) we show the proton fraction as a function of the density  for  EoSs under consideration for BEM, together with DU threshold ratios $x_{\rm
DU}(n)$. We see that the ratio $x_{\rm DU}(n)$ is only slightly model-dependent. Once the DU threshold density is reached in the center of a neutron star, the very efficient process of neutrino
cooling of a neutron star kicks in. As a result every star with a mass only slightly
exceeding $M^{(n)}_{\rm DU}$ will
be rapidly cooled down by the DU processes, $n \to p + e +\bar{\nu}_e$, $p+e\to n+\nu_e$,
even in the presence of superfluidity, and becomes almost invisible for the
thermal detection within a few years~\cite{Blaschke:2004vq,Grigorian:2005fn}.
The type II-supernova explosion scenario~\cite{Woosley} and population
synthesis models~\cite{Popov} predict that most of single neutron stars have, probably, masses below $1.5 M_{\odot}$. Therefore, one may assume that the majority of pulsars, which surface temperatures are measured, have masses $M\lsim 1.5 M_{\odot}$. The average mass
of neutron stars observed in  binaries is $1.35 M_{\odot}$. That motivated Ref.~\cite{Klahn:2006ir} to treat the inequality $M^{(n)}_{\rm DU}>1.5 M_{\odot}$ as a ``strong'' DU constraint and the inequality $M^{(n)}_{\rm DU}>1.35 M_{\odot}$ as a ``weak'' DU constraint on the nuclear EoS. The appropriate description~\cite{Blaschke:2011gc,Blaschke:2013vma} of
the data on the cooling of the pulsar in Cassiopea A
also requires absence of the DU reactions in the star interiors. (Except for the cases where one artificially increases the proton gap, which we consider as unrealistic.)
The analysis of these data
in the existing cooling scenarios supports the constraint that $M^{(n)}_{\rm DU}$ should be $\gsim 1.5 M_{\odot}$, see~\cite{Elshamouty:2013nfa,Blaschke:2013vma}. Although one can not
still exclude the possibility of lower values of $M^{(n)}_{\rm DU}$ within a more exotic explanation of the present cooling data, e.g. with artificially enhanced proton gaps, the absence of the DU process for typical neutron star configurations might be considered as the most probable scenario.

The DU thresholds and the corresponding star masses for various models are collected in Table~\ref{tab:DU-MDU}. We see that for all models, which we consider here, the values of $M^{(n)}_{\rm DU}$ prove to be fairly above the value $1.5M_{\odot}$ required by the DU constraint, although the threshold densities $n_{\rm DU}^{(n)}$ are not so large. The reason for this is that the EoSs for our models are sufficiently stiff. To show this in Table~\ref{tab:DU-MDU} we also indicate values of the maximum neutron star masses for the models under consideration.

\begin{table}
\caption{Maximum neutron star masses and DU thresholds for the KVOR, KVORcut and MKVOR models.}
 \centering
\begin{tabular}{lcccccccccc}
\hline\hline
EoS       & $M_{\rm max}~[M_\odot]$ & $n^{(n)}_{\rm DU}~[n_0]$ & $M_{\rm DU}^{(n)}~[M_\odot]$ \\
\hline
KVOR      & 2.01 & 3.96 & 1.77 \\
KVORcut04 & 2.09 & 3.38 & 1.67 \\
KVORcut03 & 2.17 & 2.85 & 1.68 \\
KVORcut02 & 2.26 & 2.43 & 1.84 \\
MKVOR     & 2.33 & 3.84 & 2.14 \\
\hline\hline
\end{tabular}
\label{tab:DU-MDU}
\end{table}

From Fig.~\ref{Pods} (left panel) we also see that in the density interval $n_0<n<4n_0$ the proton fraction of the MKVOR model is smaller than those for the KVOR and KVORcut models. For $n>4n_0$
the proton fraction for MKVOR model exceeds that for KVOR but remains smaller that those for the KVORcut models. Also we see that the smaller the value of $f_{\om}$ in KVORcut models is
chosen, the higher the proton fraction is. For $n<0.7\,n_0$ we use the BPS EoS and the curves for particle concentrations presented in Fig.~\ref{Pods} (left panel) should be replaced by that in the model for the crust.

\subsection{Gravitational mass versus baryon mass constraint}

Reference~\cite{Podsiadlowski} studying pulsar B in the double pulsar system J0737-3039 suggested
a test on the EoS of  nuclear matter. The system J0737-3039 consists of a 22.7 ms pulsar J0737-3039A, and a 2.77 ms pulsar companion J0737-3039B, orbiting near the
common center of mass in a slightly eccentric orbit of 2.4\,h duration. The gravitational mass of the pulsar B is carefully measured in~\cite{Kramer} to be $1.249 \pm 0.001M_{\odot}$.
Such a low mass could be an indication that the pulsar B was formed in a type-I supernova of an O-Ne-Mg white dwarf driven hydrostatically unstable by electron captures onto Mg and Ne.
Such an instability occurs when the progenitor star core density reaches a well-defined critical value $(\simeq 4.5\times 10^9\,{\rm g/cm^3})$ which corresponds to a well-defined mass of ONeMg core ($\simeq 1.37\,M_\odot$).
Assuming that the loss of matter during the formation of the neutron star is negligible, Ref.~\cite{Podsiadlowski} predicted the baryon mass $M_B =  u\,N_B$ to be $M_B (\mbox{J0737-3039B}) =
1.366-1.375M_{\odot}$, where $N_B$ is the total number of baryons in the star and $u$ stands for the atomic mass unit equal to $931.5$ MeV.
The baryon number is calculated as
\begin{align}
N_B = 4 \pi \int\limits_{0}^{R} \frac{dr r^2 n(r)}{\sqrt{1 - {2 G
M(r)}/{r}}},
\label{Nbar}
\end{align}
where $G$ is the gravitational constant, $R$ is the radius of the star, $n(r)$ is the baryon
number density and $M(r)$ is the total mass accumulated inside the radius $r$.

If the formation mechanism of the PSR~J0737-3039 system and the assumption of a negligible
baryon loss of companion B during its creation are valid, the result of Ref.~\cite{Podsiadlowski} provides a strong constraint on the nuclear EoS. Microscopically motivated EoSs like the relativistic Dirac-Brueckner-Hartree-Fock EoS~\cite{FuchsDBHF}, the APR EoS A18$+\delta v+$UIX$^{*}$~\cite{APR} and the AFDMC one~\cite{Gandolfi:2009nq}, as well as EoSs of many RMF-based models do not
fulfill this constraint. The baryon loss and variations of the critical mass due to carbon flashes during the collapse may result in a lowering of $M_B$ by $\lsim 1 \%\,M_\odot$, that partially
softens the constraint. However, many EoSs do not satisfy even this weaker constraint; see Ref.~\cite{Klahn:2006ir}.

In Fig.~\ref{Pods}\,(right panel) we show the gravitational neutron star mass $M_G$ versus its baryon mass $M_B$. The two empty rectangles show the change of the constraint, when the
assumed loss of the baryon number in the collapse amounts to $0.3\%M_{\odot}$ and to $1\% M_{\odot}$. Approximately the same (from $0.3\%M_{\odot}$ to $1\% M_{\odot}$) constraint box was
proposed in Ref.~\cite{Kitaura:2005bt}, which found in their model that the mass loss of the collapsing O-Ne-Mg core during the explosion leaves the neutron star with a baryon mass of $M =
1.36 \pm 0.002 M_{\odot}$. We see that the KVORcut02 EoS does not match even a weak constraint, when the assumed baryon loss in the course of the star collapse amounts to $1\% M_{\odot}$. The KVORcut04 line coincides with that of the KVOR since both EoSs differ only a little for $n<n_{\rm cen}[M_G =1.25 M_{\odot}]$ under consideration. The KVORcut03, KVOR and KVORcut04 curves match marginally the weak constraint. The MKVOR model fits marginally the ``strong'' constraint
(the curve touches the left boundary of the hatched box). Comparing the figures from the left and right panels we can find a correlation: the smaller the proton fraction is within
the density  interval $n_0<n\lsim 2.5 n_0$, the better the given EoS satisfies the
$M_G$--$M_B$ constraint. The KVORcut02 model yields the largest proton fraction in the given mass interval and does not fulfill the constraint. The proton fractions for KVOR, KVORcut04 and KVORcut03 models are almost indistinguishable  for $n\lsim 2 n_0$ and their $M_G$--$M_B$ lines almost coincide in the considered interval of masses. The MKVOR model produces the
smallest proton fraction for $n<4 n_0$ and it matches marginally
even the strong constraint.

\begin{table}
\caption{Parameters of the scaling function (\ref{etar_L}), the corresponding values of $L$ and $K_{\rm sym}$ and the central densities ($n_{\rm cen}$) of the neutron star with the mass  $1.25\,M_\odot$. }
\centering
\begin{tabular}{ccccc}
\hline\hline
$\beta$  & $\gamma$ & $L$~[MeV] & $K_{\rm sym}$~[MeV] & $n_{\rm cen}\,[n_0]$ \\
\hline
0   &   0  & 85.30 & \phantom{$-$}18.77 & 2.33 \\
0.80& 3.60 & 70.72 & $-$80.90 & 2.61 \\
0.80& 7.50 & 54.91 & $-$155.4 & 2.70 \\
1.50& 7.50 & 30.94 & $-$189.3 & 2.68 \\
\hline\hline
\end{tabular}
\label{tab:Lplay-param}
\end{table}

\begin{figure}
\centering
\includegraphics[width = 10.01cm]{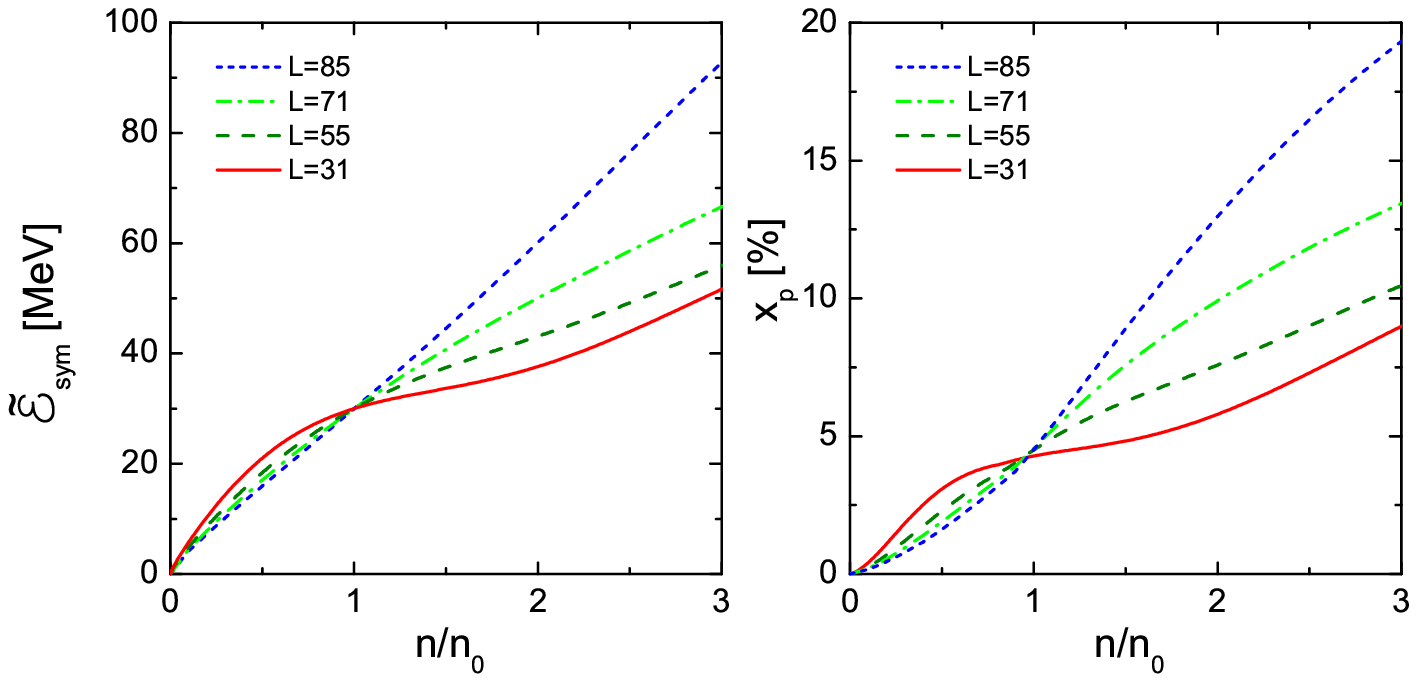}\\
\includegraphics[width = 10cm]{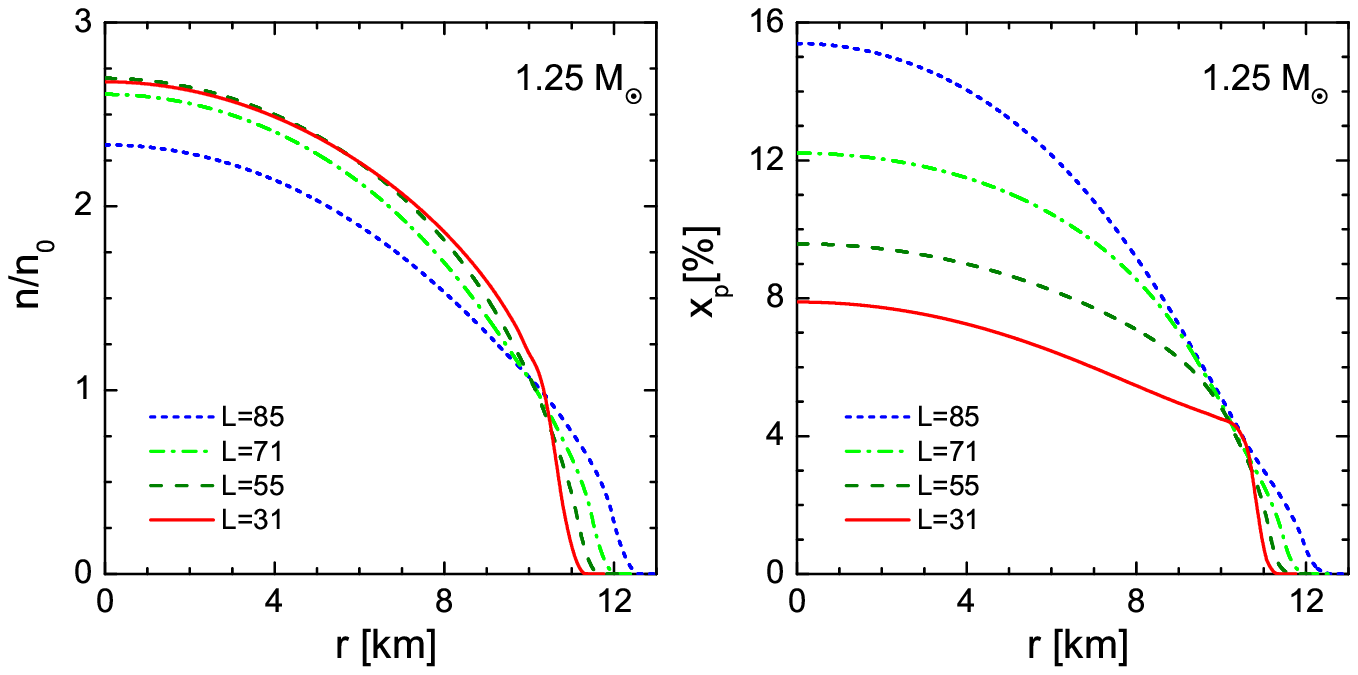}\\
\caption{Upper row: The symmetry energy (left panel) and the proton fraction (right panel) as functions of the nucleon density. Lower row: The nucleon density and the proton fraction as functions of the radial coordinate, $r$, for the neutron star with the mass $1.25\,M_\odot$. Calculations are performed in the RMF model with the parameters and the $\eta_{\sigma,\om}$ scaling functions as in the MKVOR model and the $\eta_\rho$ function given in Eq.~(\ref{etar_L}) with parameters from Table~\ref{tab:Lplay-param} for various values of $L$. The rounded values of $L$ label curves.
}\label{fig:Lplay-1}
\end{figure}

Let us elaborate more on the connection between the proton concentration and the $M_G$--$M_B$ constraint. The proton concentration is determined by the symmetry energy and therefore is correlated with the values of ${J}$ and $L$ in
Eq.~(\ref{Eexpans}). The value of ${J}$ may vary only a little from 28~MeV until 34~MeV or even in a more narrow interval, whereas the value of $L$ is less known at present~\cite{Lynch,Tsang:2012se}.
From the results shown in Fig.~\ref{Pods} we see that all KVORcut models, which miss the $M_G$--$M_B$ constraint, have a larger value of $L$ than the MKVOR model, which satisfies the constraint better.
To study how much the fulfillment of the $M_G$--$M_B$ constraint is sensitive to changes of $L$, we consider a set of test models with all parameters being the same as for the MKVOR model, except the scaling function $\eta_{\rho}(f)$ replaced by the following one
\begin{eqnarray}
\eta_\rho(f) &=& \Big(\frac{1 + \beta f}{1 + \beta \bar{f}_0}\Big)^\gamma.
\label{etar_L}
\end{eqnarray}
Varying the parameters $\beta$ and $\gamma$ we can simulate different values of $L$, see Table~\ref{tab:Lplay-param}.
In the upper row in Fig.~\ref{fig:Lplay-1} we plot the corresponding symmetry energy and the proton concentration as functions of the density. For a fixed density $n$, the decrease of $L$ in our cases leads to a decrease of  $\widetilde{\mathcal{E}}_{\rm sym}$ if  $n>n_0$ and to an increase of $\widetilde{\mathcal{E}}_{\rm sym}$ if $n<n_0$. Correspondingly, the proton concentration at a fixed density $n$ decreases with the decrease of $L$ if $n>n_0$ and increases for $n<n_0$. The EoS softens for $n>n_0$ herewith and the central density needed to support the neutron star with the fixed mass $1.25\,M_\odot$ increases from $2.33\,n_0$ to $2.70\,n_0$ for $L$ decreasing from 85~MeV to 55~MeV. Further, for $L$ decreasing from 55~MeV to 24~MeV it decreases weakly from $2.70\,n_0$ to $2.66\,n_0$. The proton concentration decreases with an $L$ decrease for all $r$, since the radius of the star decreases with  decrease of $L$.

\begin{figure}\centering
\includegraphics[width = 12cm]{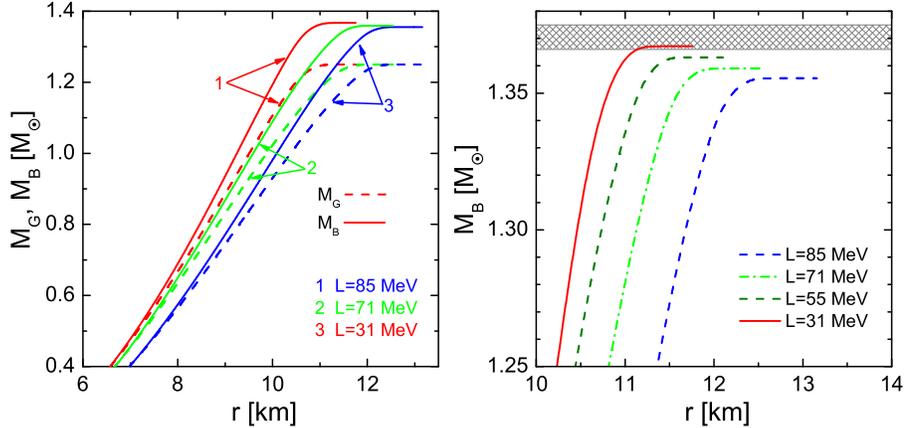}
\caption{Accumulated values of gravitational and baryon masses as functions of the radial coordinate, $r$, for various $L$. Calculations are performed for the
same models as in Fig.~\ref{fig:Lplay-1}. The double-hatched band on the right panel shows the empirical constraint from Ref.~\cite{Podsiadlowski}.
}\label{fig:Lplay-2}
\end{figure}

In Fig.~\ref{fig:Lplay-2} we depict the accumulated gravitational and baryon masses of the neutron star as functions of the radial coordinate. We observe that the baryon mass is larger than the gravitational mass, $M_{B}(r)>M_G(r)$, for any $r$. The radii at which the gravitational and baryon masses saturate, are almost the same and become smaller when the value of $L$ decreases.
The right panel in Fig.~\ref{fig:Lplay-2} demonstrates that the more we decrease the $L$ value the closer the baryon mass approaches the empirical ``strong'' constraint from~\cite{Podsiadlowski} shown by the double-hatched band, and enters it for $L\lsim 30.9$~MeV.

\subsection{Constraints on the maximum neutron star mass and radii}

Recent precise determinations of the masses of pulsar PSR~J1614-2230 in Ref.~\cite{Demorest:2010bx} as $1.97\pm 0.04 M_{\odot}$ and of pulsar PSR~J0348+0432 in Ref.~\cite{Antoniadis:2013pzd} as $2.01\pm 0.04 M_{\odot}$  delimit the EoS of neutron star matter. Other masses of the very massive pulsars are determined with large experimental error bars and with additional theoretical uncertainties.  Reference~\cite{Romani} performed initial spectroscopic observations and additional photometry from  the $\gamma$-ray  pulsar PSR J1311-3430 with the period $94$ min. Simple heated light-curve fits give the estimate for the pulsar mass $2.7M_{\odot}$. Incorporating systematic light-curve differences the authors estimated $M>2.1 M_{\odot}$.  For further information about the compact star mass-radius relation see Refs.~\cite{Lattimer:2012nd,Steiner:2012xt}.

\begin{figure}
\centering
\includegraphics[width = 10cm]{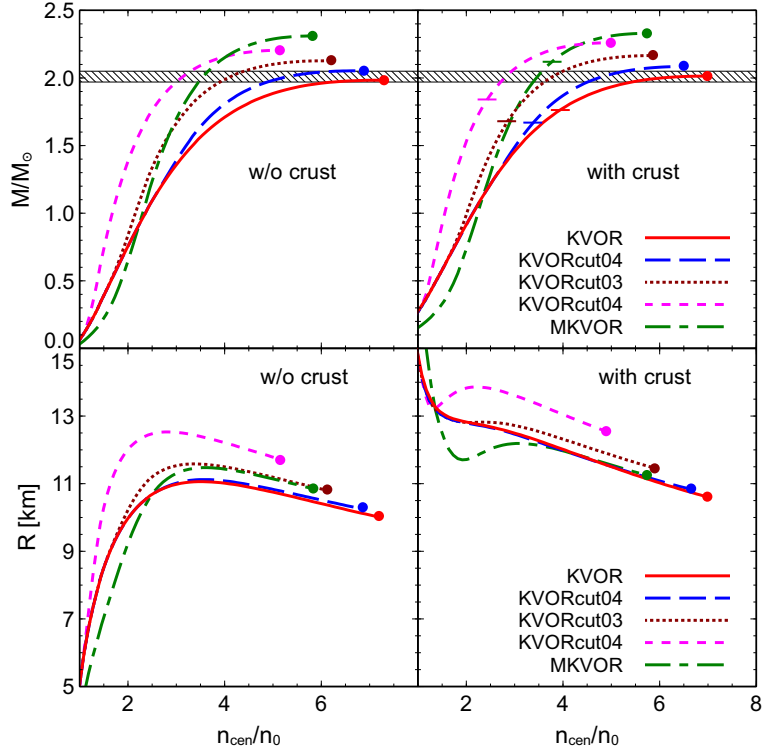}
\caption{Neutron star masses (upper panels) and radii (lower panels) versus the central density for various EoSs without (left) and with (right) inclusion of the crust.
Bold dots show maximum available masses. The band shows uncertainty
range of the mass of PSR J0348+0432 ($2.01\pm 0.04 M_{\odot}$). Horizontal dashes
in the right panel show the DU thresholds. } \label{Massdensity}
\end{figure}

In Fig.~\ref{Massdensity} we demonstrate the neutron star masses and radii as functions of the central density for our EoSs with the crust computed with the very same EoS (we name it ``without crust") and with the crust computed with BPS EoS (named "with crust").
Bold dots show maximum masses. We see that the calculation ``with crust" affects only a little
the values of the neutron star masses increasing them not more than by $0.05\,M_{\odot}$ and increasing the radii by about 1\,km for the masses corresponding to central densities 2--4\,$n_0$ and by $\sim 0.5$\,km for heavy stars with central densities 5--6\,$n_0$. The band shows the uncertainty range of the masses for PSR~J0348+0432 ($2.01\pm 0.04 M_{\odot}$). All the models fulfill the $M_{\rm max}$ constraint. We see that for the KVORcut models the lesser the value $f_{\om}$ is chosen, the larger the value of the maximum mass is and the smaller
the central density corresponding to $M_{\rm max}$ becomes. The MKVOR EoS is the stiffest among EoSs considered here. The central density $n_{\rm cen}(M_{\rm max})$  for the MKVOR EoS is above that for KVORcut02 EoS. Dashes on the right panel indicate the DU thresholds. The MKVOR model
yields the highest DU threshold mass $M^{(n)}_{\rm DU}$, cf. Table~\ref{tab:DU-MDU}.

\begin{figure}
	\centering
\includegraphics[width=9cm]{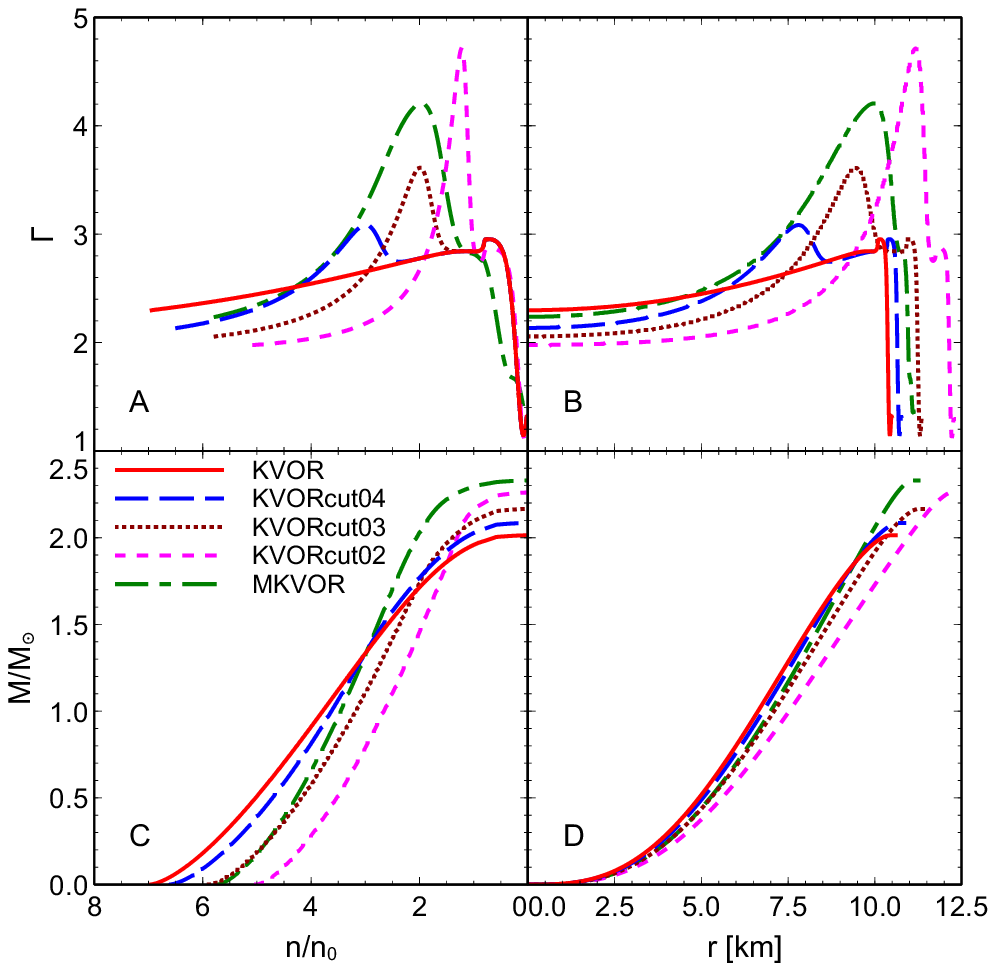}
\caption{Local stiffness (adiabatic index) of the EoS (A, B) and mass (C, D) distributions inside neutron stars with $M=M_{\rm max}$ as functions of the density $n$ (A, C) and the radial coordinate (B, D) for our EoSs. Crust is included. } \label{profiles}
\end{figure}

The internal structure of the neutron star with the maximum possible mass is illustrated in Fig.~\ref{profiles} for the various EoSs we study. In the left column we show the adiabatic index of the EoS,
\begin{align}
\Gamma(n)=\frac{n}{P}\frac{{\rm d} P}{{\rm d} n},
\label{Gamma-index}
\end{align}
as a function of $n$ varying from   the star center to the surface (A), and the corresponding part of the star mass (C) accumulated at densities smaller than given $n$.
Knowing the density profile across the star, $n(r)$, we present these quantities as functions of the radial coordinate in the right column in Fig.~\ref{profiles} (B and D). Using the adiabatic index $\Gamma$ as a measure of the local stiffness of the EoS we conclude that the $\omega$-cut procedure  stiffens strongly the EoS in comparison to the KVOR EoS in the narrow interval of densities above $n_0$. Beyond this interval at higher density the KVORcut EoSs become even softer than KVOR EoS. This can  be also observed on the accumulated mass plot. The accumulated mass $M(n)$ for the KVORcut EoS remains smaller than that for the KVOR EoS at higher $n$ inside the star and starts to exceed the mass for the KVOR EoS only in the outer part of the star. For the MKVOR EoS interval of densities, where its stiffness exceeds that of the KVOR EoS, is much broader than for any KVORcut EoS. As the result MKVOR EoS supports the neutron star with the larger maximum mass.

\begin{figure}
\centering
\includegraphics[width=12cm]{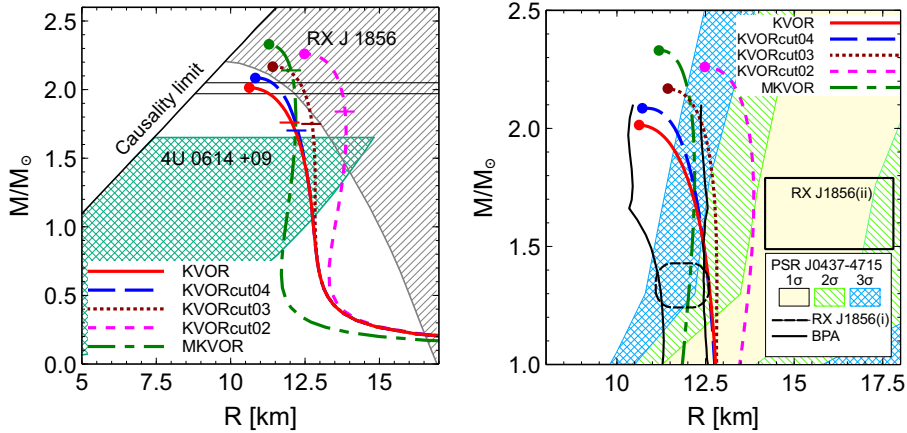}
\caption{Left panel:
Mass-radius relation for our EoSs together with constraints
from thermal radiation of the isolated neutron star RX~J1856~\cite{Trumper} and from QPOs in the LMXBs 4U 0614+09~\cite{Straaten}.  The band shows the uncertainty range of masses for
PSR~J0348+0432~\cite{Antoniadis:2013pzd}. Dashes indicate the DU thresholds.
Right panel:
Mass-radius relations for our EoSs together with other constrains on the star radius.
Hatched regions show  $1\sigma$ (light), $2\sigma$ (hatched) and
$3\sigma$ (double-hatched) confidence contours from the
mass-radius analysis~\cite{Bogdanov:2012md} of the millisecond
pulsar PSR J0437-4715. Thin lines show the $2\sigma$ contour from BPA of Ref.~\cite{Lattimer:2012nd}. The elliptic region shows the estimated $M$-$R$ constrained for RX~J1856, cf.~\cite{WCGHo}. The rectangular
box shows estimations of compactness of the isolated neutron stars
via X-ray spin phase-resolved spectroscopy from~\cite{Hambaryan2014}.
 }\label{MR}
\end{figure}

In Fig.~\ref{MR}\,(left panel) we demonstrate the
mass-radius relation for our EoSs in comparison with available experimental
constraints. In contrast to the mass determination, there are no high-accuracy
radius measurements. The highest observed quasi-periodic oscillation (QPO)
frequency (1330 Hz, for 4U 0614+091; see Ref.~\cite{Straaten})
places a constraint, see double-hatched area in the left panel.
The nearby isolated neutron star RX~J1856.5-3754
(shortly: RX~J1856) shows a purely thermal spectrum in X-rays and in
optical-UV. This allowed to determine the distance to the object~\cite{WalterLattim02} and its photospheric radius $R_\infty\sim 16.8$\,km~\cite{Trumper}.
Using the relation between the photospheric radius $R_\infty$ and the true stellar radius $R$,
$R_\infty=R(1-2GM/R)^{-1/2}$ we can express the neutron star mass as a function of the  radius as
\begin{align}
M=({2G})^{-1}{\Big(R-{R^3}/{R_\infty^2}\Big)}\,.
\end{align}
Assuming the estimate $R_\infty\gsim 16.8$\,km, which takes into account uncertainties in the distance and radius determinations, the hatched region in Fig.~\ref{MR}\,(left panel) is obtained as a constraint on the mass-radius relation~\cite{Trumper}.
There are three ways to interpret it:
(i)  RX~J1856 is a typical single star with the mass of $\simeq 1.4 M_{\odot}$, then its radius $R$ must be larger than 14\,km. Only our KVORcut02 model satisfies this constraint;
(ii) RX~J1856 has a typical radius of $R=$12--13 km, then its mass should be above $(1.8\mbox{--}2)\,M_{\odot}$, requiring  EoSs like our KVORcut03 and MKVOR EoSs;
(iii) RX~J1856 is an exotic object with a small mass $M\sim 0.2M_{\odot}$, which would be possible for all EoSs considered here.
Such small-mass neutron stars cannot be produced within standard models of the neutron star formation. Nevertheless, Ref. ~\cite{Popov04} suggested that fragmentation of rapidly rotating proto-neutron stars could lead to the formation of very light neutron stars.
In a more
detail these constraints are discussed in Ref.~\cite{Klahn:2006ir}.

In Ref.~\cite{Bogdanov:2012md} the neutron star radius is constrained to be $> 11.1$ km at $3\sigma$ confidence, assuming $M = 1.76 M_{\odot}$ for all combinations of the other
parameters, see Fig.~\ref{MR}\, (right panel). Thin lines in Fig.~\ref{MR}\, (right panel)    show the $2\sigma$ contour from  Bayesian probability
analyses (BPA) of Ref.~\cite{Lattimer:2012nd}. The elliptic region ranging from 11 km to 12.7 km shows the estimated $M$-$R$ constraint for RX~J1856, cf.~\cite{WCGHo}. We see that for all models, except KVORcut02
model, the radii of the stars lie in a narrow interval $12.5\pm
0.5$km for masses $0.5M_{\odot}\lsim M\lsim 2M_{\odot}$. This
observation puts in doubt the hope expressed in some works that a
simultaneous measurement of $M$ and $R$ may allow to fix the EoS with an appropriate certainty.
The stars computed with the KVORcut02 model yield larger radii.
The reason for this is that for the KVORcut02 model the mass and the
pressure are  higher in peripheral star regions than for other
EoSs that we consider, see Fig.~\ref{profiles}. Note that with the help of X-ray
spin phase-resolved spectroscopy, Ref.~\cite{Hambaryan2014} found
distinctly larger radii: $R > 14$~km, as is shown by the
rectangle. Among the hadron EoSs only the stiffest EoS for the KVORcut02 model may
allow to fulfill this constraint. Perhaps, such a large radii
occur for the so-called hybrid quark-hadron stars~\cite{Benic:2014jia}.
The reason why hybrid stars yield large radii is similar to that we have explained comparing our KVORcut02 EoS with other models: in hybrid star configurations one matches the hadron EoS in the shell with a softer quark one in the interior. Thus, in order to satisfy the constraint $M>2M_{\odot}$ with hybrid stars one needs to exploit  the stiffest hadron EoSs in the star shell.

Because of mentioned ambiguities of evaluations of the neutron star
radii we will consider as a relevant the
constraint of Ref.~\cite{Bogdanov:2012md}: $R> 11.1$ km at
$3\sigma$ confidence for the star with $M = 1.76 M_{\odot}$. This
constraint is well satisfied in all our models.

\section{Strangeness content. Hyperons, $\phi$ mesons, $\sigma H$ scaling.}
\label{Hyperons}

In this section we study results of the hyperonization phase
transition on the EoS within our KVOR-based models. The ratios of the
$\sigma H$ to $\sigma N$ coupling constants deduced from hyperon binding energies in ISM
following Eq.~(\ref{EHbind}) are:
\begin{align}
\mbox{KVOR, KVORcut:} \quad &
x_{\sigma \Lambda} = 0.599\,, \,\,\,
x_{\sigma \Sigma} = 0.282\,,\,\,\,
x_{\sigma \Xi} = 0.305\,,
\nonumber\\
\mbox{MKVOR:} \quad &
x_{\sigma \Lambda} = 0.607\,, \,\,\,
x_{\sigma \Sigma} = 0.378\,, \,\,\,
x_{\sigma \Xi} = 0.307\,.
\label{hyperconst}
\end{align}
The values $x_{\sigma \Lambda} =0.60$--0.61 agree well with
the best fit value derived from medium-heavy hypernuclei,
$x_{\sigma \Lambda} =0.62$, see Ref.~\cite{vanDalen:2014mqa}.

After the inclusion of hyperons we label the KVOR, KVORcut and
MKVOR models as KVORH, KVORHcut and MKVORH, respectively. Since without
hyperons KVOR and KVORcut04 models yield rather low maximum masses
and KVORcut02 model does not appropriately fulfill the flow
constraint for $n\lsim 4n_0$, we will focus our further
consideration on the best choices: KVORHcut03 and MKVORH models.
As in above consideration, all calculations below are performed
with the BPS crust EoS.

As known, the inclusion of hyperons results usually in a substantial
decrease of the maximum neutron star mass.
The difference between neutron star masses with and without hyperons proves to be so large for reasonable hyperon fractions in the standard RMF approach that in order to solve the
puzzle one needs to start with a very stiff EoS without hyperons. This hardly coincides with the predictions of the microscopic variational EoS~\cite{APR} and the quantum Monte Carlo simulations~\cite{Gandolfi:2009nq}, and is not compatible with the  constraint on stiffness of the EoS extracted  from the analysis of nucleon and kaon flows~\cite{Danielewicz:2002pu,Fuchs} in heavy-ion collision.

\begin{figure}
\centering
\includegraphics[width=11cm]{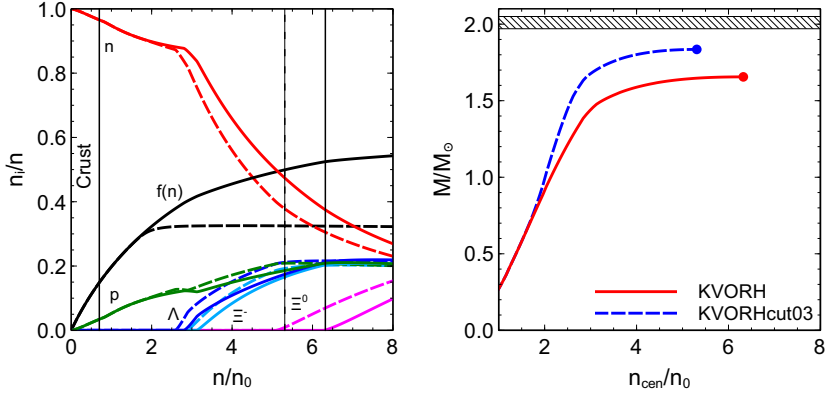}
    \caption{ Left panel: The scalar field amplitude $f(n)$, Eq.~(\ref{f-def}), and the baryon fractions as functions of the total baryon density for KVORH model (solid
        lines) and KVORHcut03 model (dashed lines) in BEM. Vertical
        thin  solid and dashed lines indicate maximally possible  values of the central
        density for  KVORH and KVORHcut03 models, respectively.  Right panel: Neutron star mass
        versus central baryon density for the KVORH model (solid line) and for the KVORHcut03 model (dashed line).
    }\label{HfracKVOR}
\end{figure}

In Fig.~\ref{HfracKVOR} (left panel) we show the dependence of the scalar field $f(n)$, see Eq.~(\ref{f-def}), and the baryon fractions on the density for KVORH model (solid lines) and
KVORHcut03 model (dashed lines) in BEM. In the right panel we
demonstrate neutron star masses versus central densities for these
models. We see that the $\Lambda$ and $\Xi^-$ begin to appear for
$n\sim (2.5\mbox{--}3)n_0$. Each of these fractions reaches   $\simeq
0.2$ for $n\gsim (5\mbox{--}6)n_0$ in both models under consideration.
The $\Xi^{0}$ hyperons appear only close to the maximum available central
density. $\Sigma^{-}$ do not appear at all. For the KVORHcut03
model the hyperon fractions prove to be larger than for the KVORH
model.

The maximum mass is $1.66 M_{\odot}$ for the KVORH model and $1.83\,M_{\odot}$ for the  KVORHcut03 model. Thus, the reductions of the maximum mass of the star equal to $0.35 M_{\odot}$  and $0.33 M_{\odot}$ for the KVORH  and KVORHcut03 model, respectively, prove to be less than those would be found in the usual RMF approach~\cite{SchaffnerBielich:2008kb,Weissenborn:2011ut}. Nevertheless, the KVORH and KVORHcut03 models with the hyperon coupling constants given by Eq.~(\ref{hyperconst}) produce too low maximum neutron star mass.
Moreover, we find that hyperons cannot be incorporated in the MKVOR model with coupling constants given by Eq.~(\ref{hyperconst}) and the  $\eta_\rho$ scaling function from~(\ref{KVORM_etar}), since the solution for $f(n)$ does not exist in this case.

\begin{figure}
\centering
\includegraphics[width=11cm]{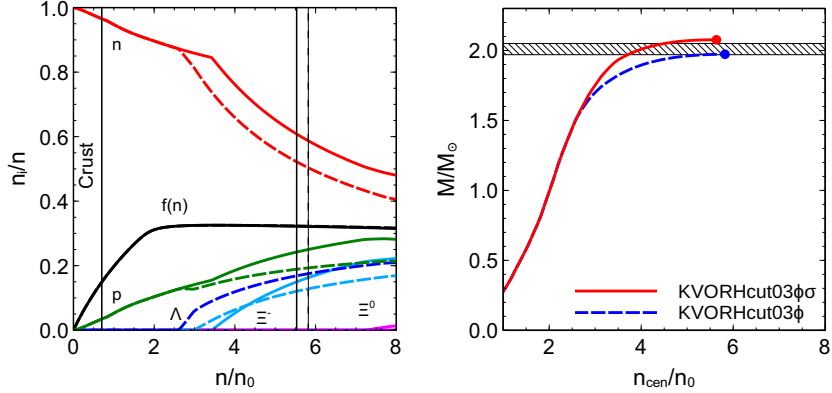}
\caption{The same as in Fig. \ref{HfracKVOR} but for models KVORH${\phi}$cut03  (dashed lines) and KVORH${\phi}{\sigma}$cut03  (solid lines).}
\label{HfracKVORHcut}
\end{figure}

To have a possibility to increase the maximum values of the neutron star masses  we incorporate the $\phi$-meson mean field and allow for a scaling of the $\sigma H$ coupling constants. We will use the very same scaling of the $\phi$-meson mass as for all other hadrons, i.e., $\Phi_{\phi}=1-f$, but with the unscaled coupling constants, $\chi_{\phi b}=1$, which gives the scaling function $\eta_{\phi}=(1-f)^2$.

First, let us use the $\phi H$ coupling constants given by Eq.~(\ref{gHm}) and the $\sigma H$ coupling constants given by Eq.~(\ref{EHbind}) with $\xi_{\sigma H}(f)=1$. So generalized KVORHcut03 and MKVORH models are labeled as the KVORHcut03$\phi$ and MKVORH$\phi$ models, respectively.

Then, additionally we incorporate $\xi_{\sigma H}\neq 1$ scaling. In order to demonstrate a qualitative effect of the $\xi_{\sigma H}$ scaling  we take $\xi_{\sigma H}(n)$ such that $\xi_{\sigma H}(n_0)=1$, cf.  Eq.~(\ref{EHbind}), and it decreases reaching
zero for the baryon density $n>n_{cH}$,  $\xi_{\sigma H}(n)=0$ for $n>n_{cH}$, where
$n_{cH}$ are critical densities for hyperonization. This assumption means that with such a parameterization we will exploit vacuum masses of the hyperon $H$ for $n>n_{cH}$. Such a models are labeled as KVORHcut03$\phi\sigma$ and MKVORH$\phi\sigma$ models.  Note that the KVOR model extended to the high temperature regime in Ref.~\cite{Khvorostukhin:2008xn} (the SHMC model) matches well the lattice data up to $T\sim 250$ MeV provided all the baryon--$\sigma$ coupling constants except the nucleon ones are artificially suppressed, that partially motivates our choice of  $\xi_{\sigma
H}$.

On the left panel in Fig.~\ref{HfracKVORHcut} we demonstrate $f(n)$ and the baryon fractions for KVORHcut03$\phi$ model (dashed lines) and for KVORHcut03$\phi\sigma$ model (solid lines). The particle fractions show a qualitatively similar behavior to that in KVORHcut03 model, except that now  the $\Xi^{0}$ do not appear at all and $\Lambda$ do not occur for KVORHcut03${\phi}{\sigma}$ model.  In the right panel we demonstrate neutron star masses as functions of the central densities for the same models. We get $M_{\rm max} {\rm [KVORHcut03\phi]}=1.97 M_{\odot}$ that marginally agrees with the observational neutron star mass constraint and $M_{\rm max} {\rm [KVORHcut03\phi\sigma]}=2.08 M_{\odot}$ that fully fulfills the maximum neutron star mass constraint.

\begin{figure}
\centering
\includegraphics[width = 11cm]{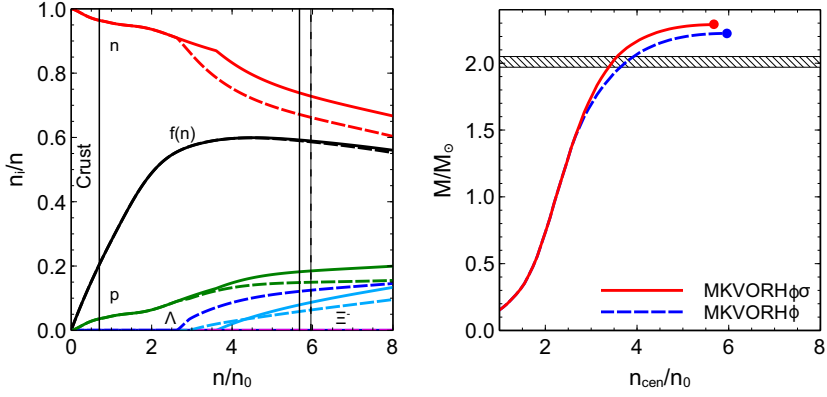}
\caption{ The same as in Fig. \ref{HfracKVOR} but for models MKVORH${\phi}$  (dashed lines) and
MKVORH${\phi}{\sigma}$  (solid lines).}
\label{ImprovedMKVORH}
\end{figure}

\begin{table}
\centering
\caption{Maximum masses, $M_{\rm max}$, and strangeness concentrations, $f_{\rm S}$ of the neutron star for various RMF models with hyperons.}
\begin{tabular}{ccccccc}
\hline \hline
& \multicolumn{2}{c}{KVOR} & \multicolumn{2}{c}{KVORcut03} &\multicolumn{2}{c}{MKVOR} \\
\cline{2-7}
& $M_{\rm max}$ [$M_\odot$] & $f_{\rm S}$ [\%] & $M_{\rm max}$ [$M_\odot$] & $f_{\rm S}$ [\%] & $M_{\rm max}$ [$M_\odot$] & $f_{\rm S}$ [\%] \\
\cline{2-7}
no hyperons               & 2.01 &0     & 2.17 & 0     & 2.33 & 0     \\
H\phantom{$\phi\sigma$}   & 1.66 &3.4   & 1.84 & 2.4   & --   & --     \\
H$\phi$\phantom{$\sigma$} & 1.88 &3.5   & 1.97 & 2.9   & 2.22 & 2.3 \\
H$\phi\sigma$             & 1.96 &0.92  & 2.08 & 0.98  & 2.29 & 0.62 \\
\hline\hline
    \end{tabular}
    \label{T_Hyp_MMax}
\end{table}

On the left panel in Fig.~\ref{ImprovedMKVORH} we show $f(n)$ and
the baryon concentrations for the MKVORH$\phi$ and MKVORH$\phi\sigma$ models.
The hyperon concentrations demonstrate a behavior similar to that in the
KVORHcut03$\phi$ and KVORHcut03$\phi\sigma$ models, being
however suppressed by a factor $\sim 0.7$. On the right panel we
demonstrate neutron star masses as functions of the central
densities for these models. $M_{\rm max} {\rm [MKVORH\phi ]}$ proves
to be $2.22 M_{\odot}$ and $M_{\rm max} {\rm [KVORHcut03\phi\sigma]}=2.29
 M_{\odot}$, both values are well above  the maximum neutron star mass constraint.

Maximum values of the masses of neutron stars and the total strangeness concentrations
calculated as the ratio of the number of strange quarks in the star to the total number of quarks given by $3N_B$, where $N_B$ is the baryon number, Eq.~(\ref{Nbar}),
\begin{align}
f_{\rm S}=\frac{1}{3N_B}\intop_0^R \frac{4\pi{\rm d} r r^2}{\sqrt{1-2GM(r)/r}} (n_\Lambda+n_{\Sigma^-}+n_{\Sigma^0}+n_{\Sigma^+}+2n_{\Xi^-}+2n_{\Xi^0})\,,
\end{align}
for the KVOR, KVORcut03 and MKVOR models and their H$\phi$ and H$\phi\sigma$ extensions
are summarized in Table \ref{T_Hyp_MMax}. We see that with inclusion of the $\phi$ meson mean-field with the scaled mass (H$\phi$ extensions of the models) the strangeness concentration increases but the maximum mass of the star increases too. Switching off the $g_{\sigma H}$ coupling constants,  as in H$\phi\sigma$ extensions, suppresses $f_{\rm S}$
drastically that leads to a further increase of $M_{\rm max}$.

\begin{table}
\centering
\caption{
Hyperon critical densities in units of $n_0$ and corresponding neutron star
masses in units of $M_{\odot}$ for KVORH, KVORHcut03 and MKVORH models.
Dashes mean that the hyperon species do not appear in the neutron star up to the maximal density.}
\begin{tabular}{l*{6}{c}}
\hline\hline
& $n_{c \Lambda}$ & $M_{\rm DU}^{(\Lambda)}$ & $n_{c \Xi^{-}}$ & $M_{c}^{(\Xi^-)}$ & $n_{c \Xi^0}$ & $M_{\rm DU}^{(\Xi^0)}$ \\
\hline
KVORH                   & 2.81 & 1.37 & 3.13 & 1.48 & 6.27 & 1.66 \\
KVORHcut03              & 2.59 & 1.51 & 2.89 & 1.65 & 5.10 & 1.84 \\
KVORHcut03$\phi$        & 2.59 & 1.51 & 2.98 & 1.69  & --   & --  \\
KVORHcut03$\phi \sigma$ & --   & --   & 3.42 & 1.91 & --   & -- \\
MKVORH$\phi$            & 2.63 & 1.43 & 2.93 & 1.65 & --   & -- \\
MKVORH$\phi \sigma$     & --   & --   & 3.61 & 2.07 & --   & -- \\
\hline\hline
\end{tabular}\label{T_Hyp_dens}
\end{table}

Hyperon critical densities and
corresponding neutron star
masses for KVORH, KVORHcut03 models and their $\phi$ and $\phi\sigma$ extensions and for extensions of the MKVORH model are presented
in Table \ref{T_Hyp_dens}. We see that before the proton
fraction reaches the DU threshold for $n\to p+e+\bar{\nu}$ reaction, the DU reactions on hyperons,
$\Lambda\to p+e+\bar{\nu}$, $p+e\to \Lambda+\nu$, and $\Xi^-\to \Lambda+e+\bar{\nu}$, $\Lambda+e\to \Xi^- +\nu$, may occur. As
we see from the Table \ref{T_Hyp_dens} in all considered
cases except KVORH and MKVORH${\phi}$, $M^{(H)}_{\rm DU}> 1.5
M_{\odot}$. Thus the KVORH and MKVORH${\phi}$ models do not
fulfill the ``strong'' DU constraint ($M_{\rm DU}>1.5 M_{\odot}$), although all models
satisfy the ``weak'' DU constraint ($M_{\rm DU}>1.35 M_{\odot}$).
After one includes in the MKVORH${\phi}$ model the $\xi_{\sigma
H}<1$ scaling the ``strong'' DU constraint can be fulfilled, as we
demonstrated on example of the MKVORH${\phi}\sigma$ model. Note that for KVORHcut03${\phi}\sigma$ and MKVORH${\phi\sigma}$ models reactions $\Xi^-\to \Lambda+e+\bar{\nu}$, $\Lambda +e\to \Xi^- +\nu$ are not allowed for
$n>n_{c \Xi^{-}}$, since there are no $\Lambda$'s, and reactions $\Xi^-\to n+e+\bar{\nu}$, $n+e\to \Xi^- +\nu$ are forbidden due to the change of the strangeness by two units. Thus, in these cases values $n_{c \Xi^{-}}$, $M_c^{\Xi^{-}}$
mean the critical density  and the mass for the appearance of $\Xi^{-}$, rather than the corresponding DU threshold values.

\begin{figure}
\centering
\includegraphics[width = 11cm]{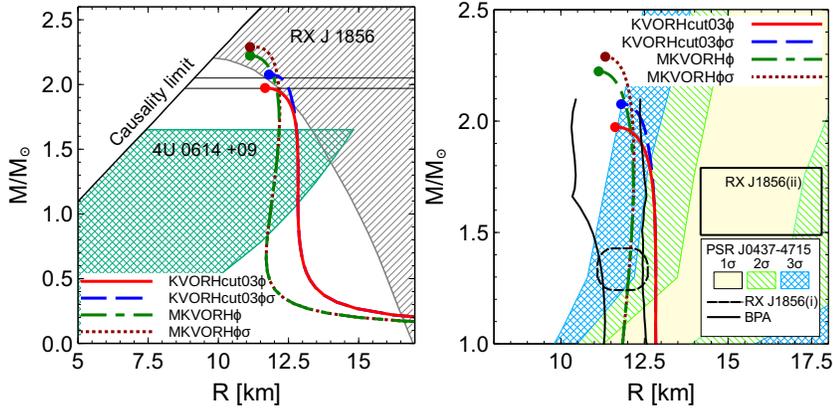}
\caption{ Mass-radius relation for KVORHcut03${\phi}$ (solid line),
 KVORHcut03${\phi}\sigma$ (dashed line), MKVORH${\phi}$ (long dash-dotted line)
 and  MKVORH${\phi\sigma}$ (dotted line) EoSs. Notations of the regions are
 the same as in Fig. \ref{MR}.
}\label{MRH}
\end{figure}

Mass-radius relations for KVORHcut03$\phi$,
KVORHcut03$\phi\sigma$, MKVORH$\phi$   and MKVORH$\phi\sigma$
EoSs are shown in Fig.~\ref{MRH}\,(left and right). We see that all presented
models satisfy the maximum neutron star mass constraint. A general
behavior of the curves  is  similar to that shown in Fig.~\ref{MR}, where
the hyperons and $\phi$ meson are not incorporated, and
$\xi_{\sigma H}=1$.

In the recent analysis~\cite{Fortin15} based on several theoretical EoSs of dense matter the conclusion was drawn that neutron stars with masses 1.0--$1.6\,M_\odot$ with necessity have radius $>13$\,km. Our cut-extensions of the KVOR model with hyperons -- KVORHcut03($\phi\sigma$) -- do agree with this conclusion rendering radii $\sim 13$\,km. However, the MKVORH($\phi\sigma$) models have smaller radii $\sim12$--12.5\,km.
The EoSs used in~\cite{Fortin15} are chosen very stiff in a pre-hyperon phase, i.e. at densities $\lsim 2$--3$n_0$, by necessity, since they have to support a sufficiently heavy neutron star even after the softening of the EoS in the hyperon phase. The stiffening of the EoS at small densities leads to an increase of the star radius as we can see at examples of the KVORcut02 and KVORcut03 models. The latter two models and the EoSs used in~\cite{Fortin15} go above the constraint~\cite{Danielewicz:2002pu} from the particle flow in HICs at least for $n\lsim 2$--3$n_0$. Differently, the MKVOR model is made softer in this density range to pass the constraint~\cite{Danielewicz:2002pu} and is stiffened at higher densities to reach a sufficiently high maximum mass of the neutron star. This makes the star radius calculated in this model larger.

\begin{figure}\centering
\includegraphics[width=5cm]{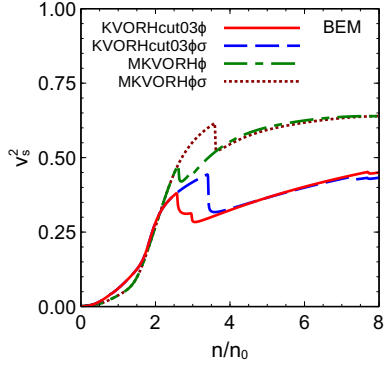}
\caption{The squared speed of sound as a function of the
baryon density for various RMF models in the BEM with hyperons.
}
\label{fig:vs-hyper}
\end{figure}

Fig.~\ref{fig:vs-hyper} presents $v_s^2 (n)$ as
a function of the baryon density for  BEM. The sharp bends appear
at $n=n_{cH}$, i.e. at the critical points of the third-order
phase transitions associated with  accumulation of the
strangeness. For $n< \mbox{min}\{n_{cH}\}$ the curves coincide with the corresponding curves in Fig. ~\ref{fig:vs}.

Thus the ``hyperon puzzle'' can be  resolved in the framework of
the models with scaled hadron masses and coupling constants provided
$\phi$ meson is included and its field mass scales at the same
rate as masses of the $\sigma$, $\omega$ and $\rho$ fields but
the $\phi H$ coupling constants  are not scaled.

\section{Conclusion}

In this paper we studied the equation of state of the cold nuclear
matter constructed within the relativistic mean-field approximation
with hadron field-mass terms and coupling constants dependent on the
$\sigma$ mean field variable $f\propto \sigma\chi_{\sigma
    N}(\sigma)$, where $\chi_{\sigma N}(\sigma)$ is the
scaling function of the $\sigma N$ coupling constant. As the simplest choice, all hadron field masses are supposed to change with $f$ at the same pace. The basics of the model were
formulated in Ref.~\cite{Kolomeitsev:2004ff} and a working model
MW(n.u., $z=0.65$) labeled in Ref.~\cite{Klahn:2006ir} as KVOR has
been constructed.  In Ref.~\cite{Klahn:2006ir} it was shown that
the KVOR model appropriately satisfies the majority of
experimental constraints known to that time. In Refs.~\cite{Khvorostukhin:2006ih,Khvorostukhin:2008xn,Khvorostukhin:2010aj,Khvorostukhin:2009pe}
the KVOR model was generalized to non-zero temperatures, the
particle excitations were included and the model was successfully
applied to the description of heavy-ion collisions. Then Refs.~\cite{Demorest:2010bx,Antoniadis:2013pzd} demonstrated new
measurements of massive neutron stars, yielding the novel
constraint on the maximum neutron star mass
$M_{\rm max}>1.97M_{\odot}$. The KVOR model, yielding $M_{\rm max}>2.01 M_{\odot}$,  fits this
constraint, although marginally.

The problem appears when one tries to include
strangeness content within the standard relativistic mean-field
framework. Owing to the hyperonization phase transition, the
equations of state are strongly  softened. Thereby it was
concluded that only  equations of state, being  extremely  stiff
without hyperons, are able to explain observed most massive
neutron stars. Inclusion of the interaction with $\phi$-meson
mean field, using artificially suppressed coupling constants, and other
modifications, as introduction of an additional repulsive terms
into the Lagrangian, although soften the problem but do not fully
solve it, see, e.g., Ref.~\cite{Weissenborn:2011ut}.Thus the aim of
the given work was to show how one can construct an appropriate set of equations of state satisfying presently known experimental constraints within the relativistic mean-field
models including hyperons and with hadron masses and coupling constants dependent of the scalar field.

First, we considered models with an artificially  suppressed
strangeness content. More specifically we focused on  two types of
modifications of the previously studied KVOR model. One type of
models (labeled KVORcut) demonstrates that the equation of state
stiffens, provided  $f(n)<1$,  if its monotonous increase for $n\sim n_{\rm tr}$ ($n_{\rm tr} >n_0$) undergoes a sharp transition to  a
constant value. The smaller a transition value of $f$ is chosen, at which the
$f(n)$ growth is quenched abruptly, the stiffer becomes the
resulting equation of state. On the example of the original Walecka model
and a non-linear Walecka model we demonstrated that suggested
$\omega$-cut scheme is applicable to all relativistic mean-field equations
of state, being a simple and efficient procedure that allows to
stiffen a given equation of state for $n>n_{\rm tr}$. In case of the KVOR-based models, this transition value of $f$ is the
smallest for KVORcut02 model and increases for KVORcut03 and then
for KVORcut04 models. The sharp change of $f(n)$ near $n_{\rm tr}$ is provided by the change of the ratio of the scalings of the hadron effective masses to the $N\omega$ meson coupling constant.

The KVOR model exploits the effective nucleon mass $m^{*}(n_0)=0.805m_N$ at nuclear saturation and incompressibility $K=275$~MeV. The new model (MKVOR) using the effective nucleon mass at
nuclear saturation $m^{*}_N (n_0)=0.73 m_N$ and incompressibility $K=240$~MeV and newly adjusted scaling functions, gives the stiffer equation of state. In the MKVOR model the sharp change of $f(n)$ near $n_{tr}$ in the beta-equilibrium neutron star matter is provided by the change of the ratio of the scalings of the hadron effective masses to the $N\rho$ meson coupling constant.

We performed a comparison with results based on the microscopic
equations of state such as the Urbana-Argonne group A18 +
$\delta v$ + UIX* one~\cite{APR}, the quantum Monte Carlo equation of
state~\cite{Lonardoni}, the Skyrme parameterizations~\cite{Cozma:2013sja}, and the chiral symmetry based calculation~\cite{Hebeler:2014ema}.
Also we calculated the Landau parameters $f_0$, $f_1$ in the isospin symmetric matter and purely neutron matter and the parameters $f'_0$, $f'_1$ in the isospin symmetric matter as functions of the baryon density within our models and compared the results with those
obtained with the Skyrme parameterizations.

Then we confronted our results to the
various constraints extracted from the numerous experimental data.
More specifically we compared the results with the nucleon optical
potential,  the heavy-ion collision nucleon and $K^{+}$ flow
constraints, the direct Urca constraint, the gravitational -- baryon
mass constraint, the maximum neutron star mass and mass-radius
constraints and with some other constraints, as
the symmetry energy constraint extracted from analysis
of the isobaric analog states,  the predictions of the
neutron-proton elliptic flow differences confronted to the
FOPI-LAND data,  the giant monopole resonance data, etc.

The most challenging is to fulfill the nucleon and kaon-flow constraints and,  simultaneously, produce a high maximum mass of a neutron star. For that we introduced the scaling functions such that our equation of state is rather soft for
$n\lsim 4 n_0$ in the isospin symmetric matter but is sufficiently stiff in the $\beta$-equilibrium matter. This behaviour is achieved by the proper selection of the scaling functions ($\eta_\om$ and $\eta_\rho$).

As an interesting finding we indicate that the gravitational -- baryon mass, $M_G$--$M_B$, constraint is
better satisfied for models with the lower proton fraction in the
density interval  $n_0 <n\lsim 2.5 n_0$. The KVORcut02 model
yields the largest proton fraction in the given density interval and does
not fulfill the constraint. The proton fractions
for KVOR, KVORcut04 and KVORcut03 models almost do not differ
for $n\lsim 2.5 n_0$ and their $M_G$--$M_B$ lines almost coincide in
the considered $M_G$--$M_B$ interval. Note that although these models
have almost the same value of the density
derivative of the symmetry energy per particle ($L/3$) as
KVORcut02 model, the proton fractions of KVOR, KVORcut03,
KVORcut04 models are essentially different from that for KVORcut02
one. The model MKVOR is the best to satisfy the $M_G$--$M_B$
constraint. In this model the proton fraction is the lowest among
considered models in the density interval  $n_0 <n\lsim 2.5 n_0$.
Also we demonstrated that the fulfillment of the $M_G$--$M_B$
constraint partially correlates with the value of $L$: the smaller
$L$ is the better the constraint is satisfied. For $L\simeq 41$
MeV even the ``strong'' constraint of Ref.~\cite{Podsiadlowski} is
fulfilled.

Another result is that the KVORcut02 model yields a larger radius of
the neutron star of the given mass than all other models
considered since the energy density and the pressure in this model are
higher at large radial coordinates, cf. Fig.~\ref{profiles}, i.e. for smaller densities $n$, than for other models. Since there are, although indirect,  indications~\cite{Bogdanov:2012md,Hambaryan2014}
 on large radii of some neutron stars ($R>14$ km), it might
indicate that the equation of state that may describe these data
should be very stiff for densities $n\lsim 4 n_0$ and it can be
softer for higher densities, e.g. such a softening might be
associated with one of possible  phase transitions at these
densities, like the hadron-quark phase transition. However we should stress that with an additional
stiffening of the equation of state for $n\lsim 4 n_0$ the
heavy-ion collision flow constraint is hardly satisfied. A weaker
constraint of~\cite{Bogdanov:2012md} that $R>11.1$ km at $3\sigma$
confidence for the star with $M=1.76 M_{\odot}$ is well satisfied
with all our equations of state.

As the next step we included hyperons. As in all known
schemes, accumulation of hyperon Fermi seas diminishes the
maximum neutron star mass. However within  our scheme this
decrease of the mass is more moderate than in the standard
relativistic mean-field approach.

Then we incorporated the $\phi$ meson mean field into consideration. We have demonstrated
that with the inclusion of the $\phi$ meson the maximum neutron star
mass increases and the corresponding KVORHcut03$\phi$ and
MKVORH$\phi$ models satisfy the maximum neutron star mass
constraint $M\geq 1.97M_{\odot}$, provided $\phi$ meson-baryon
coupling constants are unscaled, but the $\phi$ meson mass scales at the same
rate as all other hadron field masses, $M_{\rm max}[{\rm KVORHcut03\phi
}]= 1.97 M_{\odot}$ and $M_{\rm max}[{\rm MKVORH\phi}]= 2.22
M_{\odot}$. Thus, the inclusion
of the $\phi$ meson into consideration, such that the mass term scales
at the same rate as for other hadron fields and the $\phi H$
coupling constants follow the SU(6) symmetry relations, allows to solve
the part of the ``hyperon puzzle'' associated with a decrease of
the neutron star masses in the presence of hyperons. Then we demonstrated the effect on the neutron star
masses of the scaling of the $\sigma H$ coupling, $\xi_{\sigma H}(f)$, taking it such that $\xi_{\sigma H}(f(n\leq n_0))=1$ and
$\xi_{\sigma H}(f(n))$ decreases with  $n$ for $n>n_0$. As the limiting
case, we exploited the choice when the hyperon masses reach their vacuum values for $n>\mbox{min}\{n_{cH}\}$. In the corresponding models named KVORHcut03$\phi\sigma$ and MKVORH$\phi\sigma$, the maximum neutron star mass still increases, and we found $M_{\rm max}[{\rm
KVORHcut 03\phi\sigma}]= 2.07 M_{\odot}$ and $M_{\rm max}[{\rm
MKVORH\phi\sigma}]= 2.29 M_{\odot}$.

The low value of the critical density for the occurrence of $\Lambda$ hyperons in the
MKVORH$\phi$ model, $n_{c\Lambda}=2.62 n_0$,  may cause a problem with description of the neutron
star cooling  due to the opening of the efficient $\Lambda\to
p+e+\bar{\nu}$, $p+e\to \Lambda +\nu$ direct Urca reactions on the hyperon for $M>M_{\rm DU}^{(\Lambda)}=1.43\, M_{\odot}$. This part of the ``hyperon puzzle'' is avoided within the extension of the model, if one allows for the scaling of the $\sigma H$ coupling. In the MKVORH$\phi\sigma$ model $\Lambda$ hyperons and $\Xi^0$ hyperons do not appear.
Thus, we are able to resolve the
DU part of the ``hyperon puzzle''.

Other constraints considered
in the given paper are satisfied equally good in the models with
hyperons KVORHcut03$\phi$ and MKVORH$\phi$, and
KVORH\-cut03$\phi\sigma$ and MKVORH$\phi\sigma$, as it were in the
models KVORcut03 and MKVOR in the absence of hyperons.

In the given work we exploited traditional SU(6) based hyperon
couplings, whereas we could use SU(3) ones, that would allow by
varying of an extra parameter to further increase the maximum
neutron star mass. We did not include the $\Delta$ isobars into
consideration. Reference \cite{Drago:2014oja} raised a question about
the $\Delta$ puzzle. They found that for 40 MeV $<L<$ 62 MeV,
$\Delta$ isobars in beta equilibrium matter may appear at a
density of the order of $2\mbox{---}3$ times nuclear matter saturation
density, i.e. in the same range as for the appearance of hyperons.
We believe that the $\Delta$ puzzle might be solved similarly to
that for the hyperons. In a more detail this question will be
considered elsewhere.

A summary of equations of state considered in the given work in
comparison with the experimental constraints is presented in Table~\ref{T_results}. We see that KVORcut03 and MKVOR models without hyperons and KVORHcut03$\phi\sigma$, MKVORH$\phi\sigma$ models with hyperons and $\phi$ meson pass the suggested tests rather appropriately.

\section*{Acknowledgement}
This work was supported by the Ministry of Education and Science of the Russian Federation (Basic part), by the Slovak Grants No. VEGA-1/0469/15, and by ``NewCompStar'', COST
Action MP1304. Computing was partially performed in the High Performance Computing Center of the Matej Bel University using the HPC infrastructure acquired in Project ITMS 26230120002 and 26210120002 (Slovak infrastructure for high-performance computing) supported by the Research \& Development Operational Programme funded by the ERDF.

\begin{landscape}
\begin{table}[!ht]
\setlength{\tabcolsep}{7pt}
\begin{center}
  \begin{tabular}{l*{5}{c}ccccc}
        \hline\hline
       &
$\widetilde{\mathcal{E}}_{\rm sym}(n)$ &
HIC flow &
$K^+$ prod &
DU  &
\multicolumn{2}{c}{$M_G$ vs. $M_B$} &
max. mass &
\multicolumn{2}{c}{star radius}\\
&IAS & & & w./str. &1\%$M_\odot$ &0\%$M_\odot$ & $M_{\rm max}>1.97 M_\odot$ & BPA &  J0437  \\
Model  &
Fig.~\ref{fig:JLown}  &
Fig.~\ref{PressureDaniel}  &
Fig.~\ref{fig:JLown}    &
Tables \ref{tab:DU-MDU},\ref{T_Hyp_dens}&
\multicolumn{2}{c}{Fig.~\ref{Pods}} &
Figs.~\ref{MR},\ref{MRH} &
\multicolumn{2}{c}{Figs.~\ref{MR},\ref{MRH}}
 \\
 \hline
KVOR                   &$\circ$& $+$ & $+$ & $+/+$ &$\circ$& $-$   &   $+$   &$\circ$ & $+$ \\
KVORcut04              &$\circ$& $+$ & $+$ & $+/+$ &$\circ$& $-$   &   $+$   &$\circ$ & $+$ \\
KVORcut03              &$\circ$& $+$ & $+$ & $+/+$ &$\circ$& $-$   &   $+$   &$-$     & $+$ \\
KVORcut02              &$\circ$& $-$ & $-$ & $+/+$ &  $-$  & $-$   &   $+$   &$-$     & $+$ \\
MKVOR                  &  $+$  & $+$ & $+$ & $+/+$ &  $+$  &$\circ$&   $+$   &$+$     & $+$ \\
KVORHcut03$\phi$       &$\circ$& $+$ & $+$ & $+/+$ &$\circ$& $-$   & $\circ$ &$\circ$ & $+$ \\
KVORHcut03$\phi\sigma$ &$\circ$& $+$ & $+$ & $+/+$ &$\circ$& $-$   &   $+$   &$\circ$ & $+$ \\
MKVORH$\phi$           &  $+$  & $+$ & $+$ & $+/-$ &  $+$  &$\circ$&   $+$   &$+$     & $+$ \\
MKVORH$\phi\sigma$     &  $+$  & $+$ & $+$ & $+/+$ &  $+$  &$\circ$&   $+$   &$+$     & $+$ \\
\hline\hline
\end{tabular}
\end{center}
\caption{Performance of various models proposed in text with respect to constraints on the nuclear symmetry energy ($\widetilde{\mathcal{E}}_{\rm sym}(n)$) extracted from the analysis of the isobaric analog states (IAS)~\cite{DanielewiczLee}; constraints on the pressure of the nuclear matter extracted from the analysis of a particle flow in heavy-ion collisions (HIC flow)~\cite{Danielewicz:2002pu} and a $K^+$ yield in heavy-ion collisions ($K^+$ prod.)~\cite{Sagert:2011kf}; weak and strong (w./str.) constraints on the thresholds of direct Urca (DU) reactions; constraints from the relation between gravitation and baryon masses ($M_G$ vs. $M_B$) for the pulsar J0737-3039(B)~\cite{Podsiadlowski} with account for the zero-mass loss during the proto-star collapse ($0\%M_\odot$) and the loss of 1\% of the Sun mass ($1\%M_\odot$); constraint on the maximum mass of the neutron star (max. mass)~\cite{Antoniadis:2013pzd}; constraint on the neutron star radius (star radius) extracted from the Bayesian probability analysis (BPA) of several neutron stars in~\cite{Lattimer:2012nd} and from the analysis of the pulsar J0437-4715 (J0437) in~\cite{Bogdanov:2012md}.
Fulfillment (violation) of a constraint is marked by $+$($-$)  and a marginal result is marked by $\circ$.
}
    \label{T_results}
\end{table}
\end{landscape}

\appendix

\section{Matching the RMF EoS with the crust EoS}\label{app:crust}

\begin{figure}\centering
\includegraphics[width=10cm]{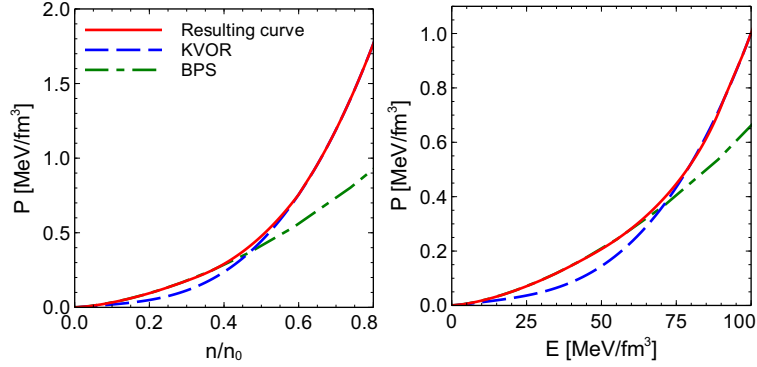}
    \caption{Pressure in the KVOR model matched with the BPS crust as a function of the number density (left panel) and as a function of the energy density (right panel). Solid line represents the resulting curve, dashed line is for the pure KVOR EoS, and long dash-dotted line is for the BPS EoS.
    }
    \label{KVOR_crust}
\end{figure}

The inclusion of the crust for all our models is performed as follows. There is always an intersection point between pressures as functions of the density in our hadron-lepton models for BEM and the BPS  EoS frequently used to describe the neutron star crust~\cite{Baym:1971pw}. For KVOR and for all KVORcut models the intersection density is $0.47 n_0$, whereas for MKVOR model it is $0.65n_0$. We have chosen a rather narrow interpolation interval for $P(E)$, and above and below it we use exactly our EoS and the BPS $P(E)$, respectively. The function $P(E)$ is what we need for integration of the Tolman-Oppenheimer-Volkoff equation to obtain the star mass  and radius, which are of our main interest (especially $M$). The interpolation interval in $E$, that we have chosen,  corresponds to the density interval $0.45 n_0 \leq n \leq 0.7 n_0$.  Interpolating EoS in this density interval, we use a cubic spline. To get thermodynamically consistent $P(n)$ and $E(n) $ dependencies we reconstruct the baryon number density $n$  as a function of $E$ using the equation
\begin{align}
\ln\frac{n_1}{n}=\intop^{n_1}_{n}\frac{{\rm d} E}{P(E)+E},
\label{eq:HP}
\end{align}
which we integrate from a density $n_1>0.7\,n_0$ downwards to lower densities. We chose $n_1$ above but not far from $0.7\,n_0$, so that at $n\sim n_1$ all KVORcut models are the same.
For $n>0.7 n_0$ we preserve our model EoSs exactly.
Here we follow the logic similar to that in Ref.~\cite{Haensel-Potekhin}. As the authors indicate there, such a procedure in not unique. However, for the present calculations, where the main ingredient for us is the function $P(E)$, it seems us optimal.
The so-reconstructed function $P(n)$ coincides with $P(n)$ from our model (KVOR, KVORcut or MKVOR)  for $n>0.7 n_0$ and agrees well (on the fraction-of-a-percent level) with $P(n)$ of the original BPS EoS for $n<0.45n_0$. Thus, the $E(n)$ dependence is thermodynamically consistent with $P(n)$ and $P(E)$ at all densities. The resulting functions $P(n)$ and $P(E)$ are shown in Fig.~\ref{KVOR_crust} for the KVOR model. The EoSs for the KVORcut02, KVORcut03, KVORcut04 models are exactly the same as for the KVOR model in the given density region.

The region of densities, where we exploit interpolating EoS, is actually the region of a pasta phase, see~\cite{Maruyama:2005vb} and references therein. The presence of the pasta affects EoS only slightly. Therefore, simplifying the consideration, we disregard this complication in this work. Note that the BPS crust EoS was also used in~\cite{Klahn:2006ir}, where it was joined with the various EoSs describing the interior region. We have also checked that a narrowing of the spline-interval  almost does not reflect on such observables, as the star mass and radius.

\section{Derivation of Landau parameters}\label{app:Landpar}

Derivations of the Landau
parameters  $f_0$ and $f_1$ are performed following the lines of  Ref. \cite{Matsui}. The parameter $f_0$ can be found with the help of our energy density functional
(\ref{Efunct}) exploiting Eqs. (\ref{NuclFL-eN}) -- (\ref{normal}).
In order to find the Landau parameter $f_1$ we need to keep the vector $\vec{\om}$ and $\vec{\rho\,}^3$ terms in the energy density.

Here we present resulting expressions for the Fermi liquid
parameters $f_0$ and $f_1$, which we calculated  for our
generalized RMF models with scaled hadron masses and couplings for ISM and PNM.

In case of the ISM ($\gamma =4$):
\begin{eqnarray}
f_0 &=& \frac{C_\om^2}{m_N^2 \eta_\om} - \frac{C_\sigma^2}{m_N^4}
\Big[ \frac{C_\om^2 n}{m_N^2 \eta_\om^2} \frac{\prt \eta_\om}{\prt
f} -\frac{m_N^2 \Phi \frac{\prt \Phi}{\prt f}}{\sqrt{p_{{\rm F}}^2
+ m_N^{*2}}}\Big]^2  \label{f0} \\ &\times& \Big((\frac{f^2}{2}\eta_\sigma
(f))''  + \frac{C_\sigma^2}{m_N^4} U''(f) + \frac{C_\sigma^2
C_\omega^2}{2 m_N^6} \Big(\frac{1}{\eta_\om(f)}\Big)'' + K_1(n, f)
+  K_2(n,f) \Big)^{-1}, \nonumber
\\ f_1 &=& -
\frac{C_\om^2}{m_N^2 \eta_\om} \frac{p_{{\rm F}}^2}{p_{{\rm
            F}}^2 + m_N^{*2}} \Big(1 +
K_3(n, f) \Big)^{-1}\,,
\label{f1}
\end{eqnarray}
\begin{eqnarray}
&&K_1(n, f) = \frac{C_\sigma^2}{m_N^2} \int\limits_0^{p_{{\rm F}}}
\frac{ \gamma p^2 dp}{2\pi^2} \frac{\Phi \frac{\prt^2 \Phi}{\prt
f^2}}{(p^2 + m_N^{*2})^{1/2}} = \frac{\gamma}{4} \frac{C_\sigma^2}{m_N^2 \pi^2}
\Phi \frac{\prt^2 \Phi}{\prt f^2} \Big( p_{\rm F} \sqrt{p_{\rm
F}^2 + m_N^{*2}} \nonumber \\ && \qquad - m_N^{*2}
\ln(\frac{p_{\rm F}}{m_N^*} + \sqrt{1 + \frac{p_{\rm
F}^2}{m_N^{*2}}}) \Big),
 \nonumber \\
&&K_2(n, f) =  \frac{C_\sigma^2}{m_N^2} \int\limits_0^{p_{{\rm
F}}} \frac{\gamma p^2 dp}{2\pi^2} \frac{(\frac{\prt \Phi}{\prt
f})^2 p^2 }{(p^2 + m_N^{*2})^{3/2}} = \frac{\gamma}{4} \frac{C_\sigma^2}{m_N^2
\pi^2} (\frac{\prt \Phi}{\prt f})^2 \Big( \frac{3 m_N^{*2} p_{\rm
F} + p_{\rm F}^3}{\sqrt{p_{\rm F}^2 + m_N^{*2}}} \nonumber \\ &&
\qquad - 3 m_N^{*2} \ln(\frac{p_{\rm F}}{m_N^*} + \sqrt{1 +
\frac{p_{\rm F}^2}{m_N^{*2}}}) \Big)\,,
 \nonumber \\
&&K_3(n,f) =  \frac{C_\om^2}{m_N^2 \eta_\om(f)}
\int\limits_0^{p_{{\rm F}}}\frac{\gamma p^2 dp}{2\pi^2}
\frac{\frac{2}{3} p^2 + m_N^{*2}}{(p^2 + m_N^{*2})^{3/2}} \nonumber \\
&& \qquad = \frac{\gamma}{2} \frac{C_\om^2}{3 m_N^2 \eta_\om(f) \pi^2} \frac{p_{{\rm
F}}^3}{\sqrt{p_{{\rm F}}^2 + m_N^{*2}}} \,,
\label{K}
\end{eqnarray}
 $f(n)$  is  solution of Eq.
\eqref{eq_fn}. Prime and double prime denote  first and  second
derivatives over $f$, respectively.

After  the  replacement
$$\frac{C_\om^2}{\eta_\om(f)} \rightarrow \frac{C_\om^2}{\eta_\om(f)} + \frac{C_\rho^2}{4 \eta_\rho(f)}\,,$$
expressions (\ref{lp_K}), (\ref{lp_ef}) and (\ref{f0}), (\ref{f1})
become valid for the PNM ($\gamma =2$).

For the isospin-dependent Landau parameters in the ISM we get:
\begin{eqnarray}
f'_0(n) &&= \frac{C_\rho^2}{4 m_N^2 \eta_\rho(f)}, \label{f0prime} \\
f'_1(n) &&= -\frac{C_\rho^2}{4m_N^2 \eta_\rho(f)} \frac{p_F^2}{p_F^2 + m_N^{*2}} \left(1 + K_4(n,f) \right)^{-1}\,,
\label{f1prime}
\end{eqnarray}

\begin{eqnarray}
K_4(n, f) &&= \frac{C_\rho^2}{4m_N^2 \eta_\rho(f)}
\int\limits_0^{p_{{\rm F}}}\frac{\gamma p^2 dp}{2\pi^2}
\frac{\frac{2}{3} p^2 + m_N^{*2}}{(p^2 + m_N^{*2})^{3/2}} \nonumber \\
&&= \frac{\gamma}{8} \frac{C_\rho^2}{3 m_N^2 \eta_\rho(f) \pi^2} \frac{p_{{\rm F}}^3}{\sqrt{p_{{\rm F}}^2 + m_N^{*2}}} \,.
\end{eqnarray}

 Another way is to find the Landau parameters $f_0, f_1$ and $f'_0$ using
 expressions (\ref{lp_K}), (\ref{lp_ef}) and \eqref{lp_j}. We have checked that
with expressions (\ref{f0}), (\ref{f1})   for $f_0$, $f_1$, and \eqref{f0prime} for $f'_0$ Eqs. (\ref{lp_K}), (\ref{lp_ef}) and \eqref{lp_j} are indeed fulfilled.


\end{document}